\documentclass[letterpaper]{mn2e}
\usepackage{graphicx,times,amsfonts}
\def \galex{{\it GALEX}}
\def \hst{{\it Hubble Space Telescope}}

\def \kms {${\rm{km}\,\rm{s}^{-1}}$}

\def \fn {$FUV-NUV$}
\def \nu {$NUV-u^*$}

\def \lgt {$\log t_{\rm SSP}$}

\def \apjs{ApJS}

\def \apj{ApJ}
\def \aj{AJ}
\def \mnras{MNRAS}
\def \araa{ARA\&A}

\title[Ultra-violet colours of passive galaxies]
{What drives the ultra-violet colours of passive galaxies?}

\author[Russell J. Smith et al.]
{Russell J. Smith$^{1}$\thanks{Email: russell.smith@durham.ac.uk}, 
John R. Lucey$^{1}$
and David Carter$^{2}$
~\\
$^1$Department of Physics, Science Laboratories, University of Durham, Durham DH1 3LE\\
$^2$Astrophysics Research Institute, Liverpool John Moores University, Twelve Quays House, Egerton Wharf, Birkenhead CH41 1LD\\
}

\date{Accepted 2012 January 9. Received 2012 January 6; in original form 2011 October 25}

\pagerange{\pageref{firstpage}$-$\pageref{lastpage}} \pubyear{2011}

\begin{document}

\label{firstpage}

\maketitle

\begin{abstract}
We present and analyse optical and ultra-violet (UV) colours for passive and optically-red Coma cluster galaxies
for which we have spectroscopic age and element abundance estimates. 
Our sample of 150 objects covers a wide range in mass, from giant ellipticals down to the bright end of 
the dwarf-galaxy regime. Galaxies with ongoing star-formation have been removed using strict H$\alpha$ 
emission line criteria. 
We focus on the colours $FUV-i$, $NUV-i$, $FUV-NUV$, $u^*-g$ and $g-i$. 
We find that all of these colours are correlated with both luminosity and velocity dispersion at the $>5\sigma$ level,
with $FUV-i$ and $FUV-NUV$ becoming bluer with increasing `mass' while the other colours become redder. 
We perform a purely empirical analysis to assess what fraction of the variation in each colour
can be accounted for by variations in the average stellar populations, as traced by the optical spectra.
For the optical colours, $u^*-g$ and $g-i$, most of the observed scatter ($\sim$80 per cent after allowing for measurement 
errors and for systematic errors in $u^*-g$) is attributable to stellar population variations, 
with colours becoming redder with increasing age and metallicity (Mg/H). 
The $FUV-i$ colour becomes bluer with increasing age and with increasing Mg/H, favouring the `metal-rich single-star' origin for the
UV upturn. However, correlations with the optically-dominant stellar populations account for only about half of the large observed scatter. 
We propose that the excess scatter in $FUV-i$ may be due to a varying proportion of ancient stars in galaxies with younger 
(SSP-equivalent) average ages. 
The $NUV-i$ colour is sensitive to SSP-equivalent age and Mg/H (in the same sense as optical colours), but also exhibits
excess scatter that can be attributed to `leakage' of the FUV-dominant old hot population. After applying a correction based on the $FUV-i$ colour, 
the much of the remaining variance in $NUV-i$ is attributable to variations in the spectroscopic parameters, similar to the results
for optical colours. 
Finally, the $FUV-NUV$ colour is surprisingly well behaved, showing strong correlations with age and metallicity, and little
residual scatter. Interpreting this colour is complicated however, since it mixes the effects of the main-sequence turn-off, in the NUV, with the 
variation in the hot post-RGB content dominating the FUV. 
\end{abstract}

\section{Introduction}

The optical and near-infrared luminosities of old ($\ga$3\,Gyr) stellar populations are dominated by 
main sequence, sub-giant and red giant branch (RGB) stars, with a combined contribution of $\sim$80\,per cent  (Maraston 2005). 
In the ultraviolet (UV) regime, by contrast,  these fairly well-constrained evolutionary phases are faint, 
and the total light output becomes sensitive to hot post-RGB phases that are much less well understood.

In particular, the spectral energy distributions of many giant ellipticals and spiral bulges exhibit an `upturn' in  the far-UV (FUV) below $\sim$2000\,\AA\
(Code \& Welch 1979). Early UV spectroscopy (e.g. Welch 1982, Ferguson et al. 1991) and imaging (e.g. Bohlin et al. 1985; O'Connell et al. 1992) 
showed that the upturn was not due to young massive stars, and hence low-mass evolved stars in old stellar 
populations are required  (see O'Connell et al. 1999 for an extensive review).
The favoured candidate sources for the UV upturn are extreme 
horizontal branch (EHB) stars, and their hot post-horizontal-branch descendants (Greggio \& Renzini 1990), which have temperatures $T_{\rm eff}$$\approx$\,25\,000\,K.
Such stars are observed in globular clusters, and also in the field, where they are designated sub-dwarf B stars 
(see Catelan et al. 2009 for a review of the EHB among other horizontal branch phenomena).
Their formation is thought to require either large mass loss on the RGB, 
occurring  at high metallicity\footnote{The formation of EHB stars in low-metallicity populations appeared unable 
to reproduce the upturn without adopting unrealistically high ages. Chung et al. (2011) have recently argued that
helium-enhanced sub-populations (for which there is independent evidence in some globular clusters) 
can develop a strong upturn at ages less than the Hubble time, since main sequence lifetimes are shorter for higher
helium abundances. 
} (Greggio \& Renzini 1990; Yi, Demarque \& Oemler 1997)
or through a variety of binary interactions (Han et al. 2007). 
An unresolved question is the degree to which these rather exotic stellar phenomena are related, 
on a broad scale, to the properties of the stellar populations which dominate the optical light of the
same galaxies. 

Observationally, one way to explore the systematic variation in FUV output for 
(optically) red galaxies is to compare UV colours against luminosity and velocity
dispersion, in analogy with the tight scaling relations obtained for optical colours. 
However, a more direct test for the physical origin of the UV flux is to compare against 
indicators for the age and metallicity of the stellar populations that dominate in the optical. Optical colours 
and individual absorption line-strengths (e.g. Mg$b$ or H$\beta$) provide some constraints
on the stellar populations, but are always sensitive to a combination of both age and metallicity effects. 
An alternative method, pursued in this paper, is to correlate the UV colours against ages, metallicities
and element abundance ratios derived from line indices, through comparison to spectral synthesis models. 
This is by no means the first attempt to address these questions, and we review some relevant previous
work in the following paragraphs. 

Using observations of 31 early-type galaxies with the {\it International Ultraviolet Explorer (IUE)},  Burstein et al. (1988) showed 
that $FUV-V$ colour becomes bluer with increasing velocity dispersion, $\sigma$, as the upturn component
strengthens in the most massive galaxies. Moreover, the same colour was found to be anti-correlated with the 
Lick Mg$_2$ absorption index. Since this index is primarily a metallicity indicator (although also sensitive to age)
Burstein et al. concluded that metallicity is the fundamental parameter underlying the observed correlations.
O'Connell (1999) repeated this test with a larger set of line indices from Trager et al. (1998), and found that 
$FUV-V$ shows clear trends with indices tracing light-element abundances, but is essentially uncorrelated with 
indices dominated by iron abundance. 

A dramatic improvement in the systematic study of UV emission from passive galaxies  
was enabled by the {\it Galaxy Evolution Explorer} (\galex) satellite (Martin et al. 2005; Morrissey et al. 2007), 
with its wide field of view and two-channel imaging capability
(NUV: $\lambda_{\rm eff}=2310\,$\AA\ FUV: $\lambda_{\rm eff}=1530\,$\AA).
In an early study of UV colours of `quiescent early-type' galaxies with \galex, Rich et al. (2005) concluded
that there was `no correlation between [the $FUV-r$ colour] and any parameter sensitive to the global metallicity', including the Mg$_2$
index used by Burstein et al. (1988). 
The Rich et al. sample was limited to luminous galaxies, with essentially no coverage at $\sigma<100$\,\kms. 
Other early \galex\ studies by Boselli et al. (2005) and Donas et al. (2007) focused on nearby galaxies, and covered a wider range in 
luminosity. Both works recovered correlations  between \fn\  and Mg$_2$, in the sense observed by Burstein et al. (1988), although 
Donas et al. show that the correlation is much weaker for $FUV-V$. They emphasise that the tight correlation in \fn\ is likely driven by metal-line
blanketing in the NUV from the main-sequence turn-off stars, rather by than variation in the hot UV-upturn sources. 

More recently, Bureau et al. (2011, hereafter B11) have revisited this issue using \galex\ colours combined with SAURON integral-field spectroscopy, 
for a sample of 48 very nearby early-type galaxies. Their analysis has an advantage over previous work, in that colours and spectroscopic 
parameters can be extracted at identical apertures. After excluding galaxies with strong H$\beta$ absorption, indicative of recent 
star-formation, B11 find that galaxies with bluer $FUV-V$ have stronger Mg$b$, similar to the original result of Burstein et al. 
(Note that B11 follow Burstein et al. in treating colour as the independent variable in the regression). 
They also find that galaxies with bluer $FUV-V$ have lower H$\beta$ absorption, suggesting older populations or higher metallicity, 
while there is no clear trend for Fe5015, an iron-dominated index. 
They conclude that $\alpha$-element abundance or total metallicity is an important driver of the UV upturn.

Although Mg$b$ (like Mg$_2$) is considered primarily a metallicity indicator, it also depends on age. Hence a correlation of $FUV-$optical
colour with this index does not unambiguously signal that the UV upturn depends on metallicity. 
Comparison of the UV colours against stellar population parameters (age, metallicity, etc) estimated from multiple indices should help
to separate the effects of the physically meaningful quantities. 
This was attempted for the $NUV-J$ colour by Rawle et al. (2008), who concluded that
metallicity effects were dominant in driving the systematic variation in this colour, with a residual dependence on age, while 
a large (0.25\,mag) excess scatter remained unexplained by the correlations. 
Loubser \& S\'anchez--Bl\'azquez (2011) performed similar comparisons for a sample of brightest cluster
galaxies (BCGs) and a control sample of similarly massive ellipticals that are not BCGs. 
They found no significant correlations of \fn\ either with linestrength indices (including Mgb5177), or with age, metallicity (Z/H) or
abundance ratio  $\alpha$/Fe taken individually. The construction of their sample, by its nature, limits the baseline in 
mass and stellar population properties which would be necessary to constrain any such correlations. Carter et al. (2011)
have applied a similar method for a sample of nearby ellipticals with spectroscopic parameters compiled from a range of literature sources. 
Testing for correlation with each single parameter in turn, they found significant negative correlations of \fn\ with total metallicity (Z/H), and 
with $\alpha$/Fe but not with iron abundance (Fe/H); the age dependence was not explored. 

In this paper, we use deep \galex\ imaging of the Coma cluster, together with optical data from the 
Canada--France--Hawaii Telescope (CFHT) and the Sloan Digital Sky Survey (SDSS), to measure ultraviolet and optical colours for a sample of 
red-sequence galaxies for which ages, metallicities and abundance ratios have been determined from high-S/N spectroscopy. 
The key improvements over previous works are (1) the use of (in principle) meaningful stellar population
parameters, rather than individual indices which mix age and metallicity effects, and (2) application to a sample spanning
a wide range in galaxy mass ($\sigma=50-400$\,\kms).
We analyse the correlations of the colours with spectroscopic ages and abundances, making use of multi-parametric fits to disentangle the
effects of each parameter. In particular, we aim to determine what fraction of the variation in each colour is driven by 
variations in the stellar populations, as constrained from the optical, and what fraction is `excess scatter', apparently unrelated to the 
optically-dominant populations. Our approach is purely empirical, and independent of any population synthesis models for the UV colours.

The structure of the paper is as follows: the optical spectroscopy and the \galex\ and optical imaging data are described in Section~\ref{sec:data}.
The results from fitting the colours with single- and multi-parameter models are presented in Section~\ref{sec:results}, together with an exploration
of residual trends and tests for effects of systematic errors.  We discuss the implications of our results  in Section~\ref{sec:discuss},
focusing especially on the questions of the origin of the UV upturn, and whether the scatter in the NUV implies widespread recent star formation 
among optically passive galaxies. Our conclusions are summarised in Section~\ref{sec:concs}.

\section{Data}\label{sec:data}

Our approach is to measure UV/optical colours for galaxies drawn from a catalogue of galaxies with 
measured spectroscopic ages and metallicities. 
Full details of the spectroscopy are given in related papers (Smith et al. 2012; Smith et al., in preparation). 
Broadly the parent sample comprises spectra for 242 bright Coma cluster members from SDSS (analysed in Price et al. 2010)
and 169 fainter members observed using long integrations  with Hectospec at the 6.5m MMT. 
The MMT sample is an extended version of that reported by Smith et al. (2009). 
Velocity dispersions were compiled from a variety of literature sources, supplemented with new observations from VLT/FLAMES, and combined
using observations from multiple data-sources to determine relative systematic offsets. 

The combined MMT and SDSS sample has been homogeneously re-analysed, to measure
absorption line strength indices and emission line equivalent widths (used in defining a passive galaxy sample). 
Ages, metallicities (Fe/H), and abundance ratios (Mg/Fe, Ca/Fe, C/Fe and N/Fe) were measured
via comparison to the Schiavon  (2007) simple stellar population (SSP) models, using a new model inversion code.
The first step in this process is a non-linear optimisation to derive Fe/H and age assuming solar abundance ratios, 
using the observed H$\beta$ and $\langle$Fe$\rangle$ = $\frac{1}{2}$(Fe5270+Fe5335) indices. 
Then we use a fast linear fit (i.e. a matrix inversion) to estimate a correction to the abundance pattern
required to match additional indices (Mg$b$, Ca4227, C$_2$4668 and CN$_2$), at the fitted age and Fe/H.  
Using this updated abundance pattern, we repeat the non-linear fit to obtain an improved estimate of the age and Fe/H. 
Alternating between these steps, the process converges within a few iterations. Parameter errors (and covariances) are obtained by propagating the index errors 
through the fitting process using montecarlo simulations (50 realisations per galaxy). We allow extrapolation of predictions beyond the model limits (e.g. to unphysically
large ages), in order to characterise the errors correctly. 
We have confirmed that our method yields results consistent with the much slower method of Graves \& Schiavon (2008), when applied to
the same data. The derived properties (especially age) should be interpreted as `SSP-equivalent parameters', i.e. they are the parameters
of the single-burst, single-metallicity, single-abundance-mixture population which best reproduces the measured indices.

We draw the UV data from proprietary and archival \galex\ observations in NUV and FUV. 
There are two deep observations: GI5\_025001\_COMA with 15\,ksec integration in the cluster core (see also Smith et al. 2010) 
and GI2\_046001\_COMA3 with 30\,ksec exposure in a field centred 0.9\,degrees to the south-west (see also Hammer et al. 2010). Retrieving additional
archival tiles with integration times 1--5\,ksec\ from the archive resulted in coverage for almost all galaxies in our spectroscopic sample.
For galaxies covered by more than one \galex\ tile, we employ only the deepest available observation.
Deep observations with \galex\ are confusion-limited, and many of the faintest objects in our sample would not be individually detected by 
running, e.g. SExtractor (Bertin \& Arnouts 1996) on the \galex\ images. Instead, we simply extract the UV fluxes from circular apertures
with centres fixed at the optical coordinates. We adopt an aperture of 10\,arcsec diameter, corresponding to $\sim$5\,kpc at the distance of Coma, 
which is sufficiently large to avoid sensitivity to variation in the \galex\ point-spread function (5--6\,arcsec FWHM). 
A local sky estimate is obtained from an annulus of radius 30--60\,arcsec, with contaminating sources identified by k-sigma clipping
the (deeper) NUV images, and subsequently excised from the sky statistics in both bands.

For optical photometry in redder passbands, we use aperture magnitudes from the 7th Data Release of the Sloan Digital Sky Survey, 
for the $g$ and $i$ bands, interpolating from the published apertures to the 10\,arcsec diameter used for the UV. 
Since the SDSS $u$-band photometry is very shallow, we instead use dedicated observations obtained with MegaCam at the 
Canada--France--Hawaii Telescope (CFHT). The CFHT $u^*$-band imaging covers a 2.8$\times$2.8\,deg$^{2}$ area with a grid of nine 
slightly-overlapping pointings. The  $u^*$ filter is $\sim$200\,\AA\ redder than SDSS $u$. 
The individual exposures for each $\sim$1\,deg$^2$ pointing were processed and stacked at the observatory using the Elixir pipeline (Magnier \& Cuillandre 2004).
Magnitudes in 10-arcsec apertures were measured from the stacks, which  have a total exposure time of 1360\,sec.

The adopted random errors in the aperture magnitudes and colours include contributions from photon statistics and from confusion noise.
We estimate the latter in all five bands, by measuring the 68 percentile interval of the total flux in 10-arcsec apertures placed randomly in a 
representative image, and propagating the effects into errors as a function of galaxy magnitude. 
Confusion noise is the dominant  source of random error in the UV for the deepest GALEX tiles, 
due to the low spatial resolution of the images, and also because the cluster galaxies are relatively faint in the UV compared to the background sources. 
Even in the optical, however, confusion can be comparable to the (generally very small) photon errors, given the large apertures used for the photometry. 
The adopted errors do not include contributions from systematic calibration error, since many of the galaxies in the sample are drawn from only a small number
of observations (i.e. GALEX tiles, SDSS drift-scan runs, MegaCam pointings) per data source, and hence calibration errors affect large parts of the sample
coherently. We assess the impact of systematic calibration errors in  Section~\ref{sec:calerrs} by explicitly allowing additional `offset' terms in our analysis. 

No corrections for galactic extinction are applied. At the high galactic latitude of Coma, the mean reddening from Schlegel et al. (1998)
is $\langle E(B-V)\rangle$=0.01\,mag, with a galaxy-to-galaxy scatter of 0.001\,mag. Even in the NUV (the band most sensitive to extinction), 
this would introduce a scatter of $<0.01$\,mag in flux, which is negligible in comparison to the observed scatter ($\sim$0.2\,mag).
No $k$-corrections are applied, since they are necessarily model-dependent and very uncertain in the UV bands. 
Although all sample galaxies lie at approximately the same distance, their recession velocities span a range 3800--9400\,\kms. Hence 
band shifting effects may be relevant, for example in the $u^*$ band, where the long-wavelength filter cut-off is close to the steep 4000\,\AA\ break. 
We address this point in Section~\ref{sec:kcorr} by testing fits which explicitly include a redshift-dependent term.

From the full spectroscopic dataset, we restrict the sample to galaxies having age greater than 2\,Gyr, error less than 0.3 in \lgt\ and [Mg/H], and
FUV error less than 0.3\,mag. 
Furthermore, we retain only  galaxies with measured velocity dispersions, and 
with $<$0.5\,\AA\ equivalent width in H$\alpha$ emission (after separation from the absorption line).
The purpose of the H$\alpha$ cut is primarily to remove galaxies in which the  stellar H$\beta$ absorption line, used in the estimation of ages and metallicities, is 
contaminated by nebular emission. In addition, the selection against H$\alpha$ emission helps to remove objects whose UV colours are influenced
by ongoing star formation in their central regions. (We address the possibility of star formation beyond the spectroscopic fibre in Section~\ref{sec:mismatch}.)
We automatically reject galaxies with UV flux centroids offset from the nominal galaxy position by more than 1.5\,arcsec, to select against
objects contaminated by neighbouring sources. (The  centroid shift threshold is supported by simulations of isolated poisson-sampled PSFs, 
tuned to the depth of the shallower tiles). Finally, all remaining sources were inspected visually on the \galex\ and optical images, 
and five galaxies manually removed from the sample due to blending undetected by the centroid offset, or location within the halo of a UV-bright star.

The resulting sample comprises 150 galaxies. For 90 objects (60 per cent), the UV measurements are from the deep Coma Core tile, while
for 25 (17 per cent), the UV data are from the deep Coma South West tile. The remaining 35 objects (23 per cent) are from the shallower tiles. 
The merged sample is mainly (97 per cent) comprised of galaxies with SDSS spectroscopy, 
since the UV flux errors are typically too large for the fainter galaxies observed that were observed with MMT. 
The final dataset used for the analysis in this paper is reported in Tables~\ref{tab:uvdattab1}--\ref{tab:uvdattab3}.
Note that the derived parameters for SDSS galaxies differ slightly from those tabulated in Price et al. (2010) due to different 
conventions and methods adopted in reanalysing the index data.

\begin{table*}
\caption{The galaxy properties used for the analysis in Section~\ref{sec:results}.
The identifications are from Godwin, Metcalfe \& Peach (1983), where available, with coordinates derived from matching to SDSS. 
The parameters used as predictors for the colours are the r-band luminosity from SDSS ($L_{r}$), the velocity dispersion ($\sigma$) from the 
complilation in Smith et al. (2012), and the SSP-equivalent age and metallicity ($T_{\rm SSP}$ and [Mg/H]) from Smith et al. (2012). 
The errors on the latter parameters are highly anti-correlated. The error coefficient $\rho_{\rm err}$ is needed when computing errors on
combinations of age and metallicity.  
The complete contents of this table are provided in the online version of the journal. 
 }
\label{tab:uvdattab1}
\begin{tabular}{lccccccc}
\hline
Galaxy & 
\multicolumn{1}{c}{R.A.} & 
\multicolumn{1}{c}{Dec} & 
\multicolumn{1}{c}{$\log(L_r/L_\odot)$} & 
\multicolumn{1}{c}{$\log(\sigma / {\rm km\,s^{-1})}$} &
\multicolumn{1}{c}{$\log(T_{\rm SSP} / 10\,{\rm Gyr})$} &
\multicolumn{1}{c}{[Mg/H]} & 
\multicolumn{1}{c}{$\rho_{\rm err}$} \\
\hline
GMP6617 & 12:54:05.52 & +27:04:07.0 & 10.16 & $2.065\pm0.015$ & $ 0.920\pm0.151$ & $\phantom{+} 0.010\pm0.109$ & $-0.86$ \\
GMP6568 & 12:54:09.98 & +28:05:33.0 & 10.13 & $2.223\pm0.011$ & $ 0.948\pm0.115$ & $\phantom{+} 0.239\pm0.077$ & $-0.79$ \\
GMP6545 & 12:54:16.03 & +27:18:13.5 & 10.26 & $2.213\pm0.014$ & $ 0.737\pm0.162$ & $\phantom{+} 0.261\pm0.116$ & $-0.92$ \\
GMP6503 & 12:54:22.28 & +27:05:03.0 & \phantom{1}9.77 & $2.158\pm0.013$ & $ 1.012\pm0.146$ & $\phantom{+} 0.236\pm0.107$ & $-0.88$ \\
GMP6404 & 12:54:32.96 & +28:22:36.4 & 10.24 & $2.131\pm0.014$ & $ 0.877\pm0.158$ & $\phantom{+} 0.242\pm0.095$ & $-0.88$ \\
GMP6409 & 12:54:36.82 & +26:56:05.5 & 10.06 & $1.868\pm0.029$ & $ 0.581\pm0.181$ & $\phantom{+} 0.096\pm0.150$ & $-0.88$ \\
GMP5978 & 12:55:27.79 & +27:39:22.0 & 10.25 & $2.191\pm0.013$ & $ 0.893\pm0.128$ & $\phantom{+} 0.219\pm0.090$ & $-0.81$ \\
GMP5975 & 12:55:29.10 & +27:31:17.2 & 10.51 & $2.275\pm0.004$ & $ 0.810\pm0.142$ & $\phantom{+} 0.321\pm0.079$ & $-0.88$ \\
GMP5886 & 12:55:41.30 & +27:15:02.7 & 10.36 & $2.408\pm0.010$ & $ 1.099\pm0.110$ & $\phantom{+} 0.372\pm0.083$ & $-0.84$ \\
GMP5704 & 12:56:01.75 & +26:45:23.8 & \phantom{1}9.78 & $2.195\pm0.015$ & $ 1.044\pm0.263$ & $\phantom{+} 0.187\pm0.140$ & $-0.91$ \\
GMP5641 & 12:56:06.40 & +27:38:52.0 & \phantom{1}9.69 & $1.828\pm0.034$ & $ 0.330\pm0.043$ & $\phantom{+} 0.184\pm0.039$ & $-0.74$ \\
GMP5599 & 12:56:09.90 & +27:50:39.3 & 10.04 & $2.088\pm0.013$ & $ 0.801\pm0.132$ & $\phantom{+} 0.313\pm0.081$ & $-0.87$ \\
GMP5526 & 12:56:16.69 & +27:26:45.5 & \phantom{1}9.82 & $2.119\pm0.013$ & $ 0.991\pm0.277$ & $\phantom{+} 0.199\pm0.179$ & $-0.94$ \\
GMP5495 & 12:56:19.79 & +27:45:03.7 & 10.09 & $2.267\pm0.013$ & $ 1.046\pm0.172$ & $\phantom{+} 0.381\pm0.113$ & $-0.88$ \\
GMP5428 & 12:56:26.63 & +27:49:50.3 & 10.11 & $2.178\pm0.010$ & $ 1.011\pm0.129$ & $\phantom{+} 0.324\pm0.096$ & $-0.83$ \\
GMP5424 & 12:56:29.13 & +26:57:25.2 & 10.29 & $2.204\pm0.012$ & $ 0.884\pm0.156$ & $\phantom{+} 0.158\pm0.094$ & $-0.90$ \\
GMP5397 & 12:56:29.82 & +27:56:24.0 & 10.08 & $2.253\pm0.010$ & $ 1.089\pm0.133$ & $\phantom{+} 0.302\pm0.100$ & $-0.87$ \\
GMP5395 & 12:56:32.04 & +27:03:20.1 & \phantom{1}9.91 & $2.212\pm0.010$ & $ 0.935\pm0.168$ & $\phantom{+} 0.264\pm0.094$ & $-0.88$ \\
GMP5364 & 12:56:34.19 & +27:32:20.2 & 10.02 & $2.197\pm0.009$ & $ 0.848\pm0.142$ & $\phantom{+} 0.361\pm0.103$ & $-0.83$ \\
GMP5341 & 12:56:35.20 & +28:16:31.6 & \phantom{1}9.89 & $2.109\pm0.016$ & $ 0.766\pm0.166$ & $\phantom{+} 0.267\pm0.110$ & $-0.84$ \\
GMP5272 & 12:56:42.86 & +28:01:13.6 & 10.29 & $2.260\pm0.010$ & $ 1.023\pm0.131$ & $\phantom{+} 0.339\pm0.093$ & $-0.86$ \\
GMP5283 & 12:56:43.52 & +27:02:05.1 & \phantom{1}9.64 & $1.891\pm0.029$ & $ 0.799\pm0.241$ & $\phantom{+} 0.030\pm0.142$ & $-0.85$ \\
GMP5279 & 12:56:43.52 & +27:10:43.7 & 10.65 & $2.374\pm0.003$ & $ 1.102\pm0.132$ & $\phantom{+} 0.311\pm0.085$ & $-0.87$ \\
GMP5250 & 12:56:47.77 & +27:25:15.6 & \phantom{1}9.56 & $1.797\pm0.027$ & $ 0.510\pm0.170$ & $\phantom{+} 0.126\pm0.137$ & $-0.72$ \\
GMP5191 & 12:56:53.14 & +27:55:46.2 & \phantom{1}9.79 & $2.319\pm0.013$ & $ 1.247\pm0.190$ & $\phantom{+} 0.281\pm0.138$ & $-0.82$ \\

\hline
\end{tabular}
\end{table*}

\begin{table*}
\caption{The 10-\,arcsec diameter aperture magnitudes from which colours were derived for the analysis in Section~\ref{sec:results}.
The  $FUV$ and $NUV$ magnitudes are from {\it GALEX}, $g$ and $i$ are from SDSS (interpolated from DR7 aperture photometry), 
and $u^*$ is from CFHT (this passband is centred $\sim$200\,\AA\ redder 
than SDSS $u$). All magnitudes are on the AB system. The quoted errors include a contribution from confusion noise, but not for calibration uncertainty.
Neither extinction corrections nor $k$-corrections have been applied.
The complete contents of this table are provided in the online version of the journal. 
}
\label{tab:uvdattab3}
\begin{tabular}{llrrrrr}
\hline
Galaxy & 
\multicolumn{1}{l}{GALEX tile} & 
\multicolumn{1}{c}{$FUV_{10^{\prime\prime}}$} & 
\multicolumn{1}{c}{$NUV_{10^{\prime\prime}}$} & 
\multicolumn{1}{c}{$u^*_{10^{\prime\prime}}$} & 
\multicolumn{1}{c}{$g_{10^{\prime\prime}}$} & 
\multicolumn{1}{c}{$i_{10^{\prime\prime}}$} \\
\hline
GMP6617 & NGA\_DDO154 & $22.525\pm0.119$ & $21.094\pm0.043$ & $17.257\pm0.003$ & $15.891\pm0.002$ & $14.724\pm0.004$ \\
GMP6568 & GI1\_039003\_Coma\_MOS03 & $22.141\pm0.137$ & $21.180\pm0.052$ & $17.225\pm0.003$ & $15.747\pm0.002$ & $14.545\pm0.002$ \\
GMP6545 & NGA\_DDO154 & $21.736\pm0.074$ & $20.701\pm0.032$ & $16.991\pm0.002$ & $15.590\pm0.002$ & $14.390\pm0.001$ \\
GMP6503 & NGA\_DDO154 & $22.903\pm0.153$ & $21.695\pm0.068$ & $17.754\pm0.005$ & $16.323\pm0.003$ & $15.116\pm0.002$ \\
GMP6404 & GI1\_039003\_Coma\_MOS03 & $22.308\pm0.152$ & $21.388\pm0.061$ & $17.444\pm0.004$ & $15.956\pm0.002$ & $14.744\pm0.002$ \\
GMP6409 & NGA\_DDO154 & $23.203\pm0.187$ & $21.403\pm0.054$ & $17.629\pm0.004$ & $16.389\pm0.003$ & $15.315\pm0.003$ \\
GMP5978 & COMA\_SPEC\_A & $22.325\pm0.171$ & $21.090\pm0.048$ & $17.129\pm0.003$ & $15.640\pm0.002$ & $14.435\pm0.001$ \\
GMP5975 & GI2\_046001\_COMA3 & $21.710\pm0.038$ & $20.508\pm0.020$ & $16.382\pm0.001$ & $14.949\pm0.011$ & $13.742\pm0.009$ \\
GMP5886 & GI2\_046001\_COMA3 & $20.551\pm0.017$ & $20.169\pm0.015$ & $16.623\pm0.002$ & $15.184\pm0.001$ & $13.943\pm0.001$ \\
GMP5704 & GI2\_046001\_COMA3 & $22.544\pm0.072$ & $21.464\pm0.046$ & $17.595\pm0.004$ & $16.278\pm0.003$ & $15.095\pm0.002$ \\
GMP5641 & GI2\_046001\_COMA3 & $23.742\pm0.194$ & $21.759\pm0.059$ & $17.922\pm0.005$ & $16.613\pm0.004$ & $15.591\pm0.004$ \\
GMP5599 & COMA\_SPEC\_A & $22.996\pm0.261$ & $21.513\pm0.065$ & $17.744\pm0.005$ & $16.233\pm0.003$ & $15.014\pm0.002$ \\
GMP5526 & GI2\_046001\_COMA3 & $23.188\pm0.121$ & $21.740\pm0.058$ & $17.824\pm0.005$ & $16.474\pm0.003$ & $15.291\pm0.003$ \\
GMP5495 & COMA\_SPEC\_A & $22.064\pm0.150$ & $21.034\pm0.046$ & $17.198\pm0.003$ & $15.659\pm0.002$ & $14.432\pm0.001$ \\
GMP5428 & COMA\_SPEC\_A & $22.301\pm0.169$ & $21.048\pm0.047$ & $17.273\pm0.003$ & $15.788\pm0.002$ & $14.590\pm0.002$ \\
GMP5424 & GI2\_046001\_COMA3 & $22.046\pm0.049$ & $20.637\pm0.022$ & $16.757\pm0.002$ & $15.405\pm0.002$ & $14.221\pm0.001$ \\
GMP5397 & GI1\_039003\_Coma\_MOS03 & $21.967\pm0.125$ & $21.165\pm0.052$ & $17.442\pm0.003$ & $15.960\pm0.003$ & $14.762\pm0.003$ \\
GMP5395 & GI2\_046001\_COMA3 & $22.681\pm0.080$ & $21.068\pm0.032$ & $17.207\pm0.003$ & $15.881\pm0.002$ & $14.741\pm0.002$ \\
GMP5364 & GI2\_046001\_COMA3 & $22.278\pm0.058$ & $21.335\pm0.041$ & $17.411\pm0.003$ & $15.980\pm0.002$ & $14.745\pm0.002$ \\
GMP5341 & GI1\_039003\_Coma\_MOS03 & $23.093\pm0.241$ & $21.582\pm0.070$ & $17.626\pm0.004$ & $16.208\pm0.003$ & $15.031\pm0.002$ \\
GMP5272 & GI1\_039003\_Coma\_MOS03 & $21.918\pm0.122$ & $20.959\pm0.045$ & $17.072\pm0.002$ & $15.575\pm0.002$ & $14.343\pm0.001$ \\
GMP5283 & GI2\_046001\_COMA3 & $23.728\pm0.191$ & $21.535\pm0.049$ & $17.837\pm0.005$ & $16.603\pm0.004$ & $15.519\pm0.003$ \\
GMP5279 & GI2\_046001\_COMA3 & $20.808\pm0.020$ & $20.120\pm0.015$ & $16.428\pm0.001$ & $14.952\pm0.001$ & $13.693\pm0.001$ \\
GMP5250 & GI2\_046001\_COMA3 & $23.960\pm0.233$ & $21.833\pm0.063$ & $18.136\pm0.007$ & $16.954\pm0.005$ & $15.928\pm0.005$ \\
GMP5191 & COMA\_SPEC\_A & $22.263\pm0.167$ & $21.278\pm0.055$ & $17.552\pm0.004$ & $16.105\pm0.003$ & $14.899\pm0.002$ \\

\hline
\end{tabular}
\end{table*}

\section{Linear analysis of the colours}\label{sec:results}

\subsection{Methodology}

In this paper we take a purely empirical and statistical approach to assessing the extent to which variations in various physical parameters 
can account for the variation in the UV and optical colours of non-star-forming galaxies. We make no use of population synthesis models to predict the colours. 
Instead, we perform linear regression analyses, seeking to find the predictor variable (or variables) 
which best reproduce the observed colours. 

We restrict attention to a subset of colours chosen to highlight the behaviour in the UV in comparison to the optical, with little redundancy:
\begin{enumerate}
\item $FUV-i$ :  directly probes the UV-upturn sources, measuring the flux from old hot stars versus that from old cool stars.
\item $NUV-i$ :  the NUV flux likely has contributions from the MS turn-off as well as from the upturn sources.
\item $FUV - NUV$ :  a colour measurable from {\it GALEX} alone. 
\item $u^* - g$ : a blue-optical colour, measuring the 4000\,\AA\ break, to test the transition between UV and optical regimes.
\item $g - i$ : a broad optical colour for comparison.
\end{enumerate}

We fit the observed colours to the following linear models in which the predicted colour is $p$, given by: 
\[p = a_{0,L} + a_{L} \log(L_r/10^{10}L_{r,\odot})\]
\[p = a_{0,\sigma} + a_{\sigma} \log(\sigma / 100\,{\rm km\,s^{-1}})\]
\[p = a_{0,t} + a_{t} \log(T_{\rm SSP}/10\,{\rm Gyr})\]
\[p = a_{0,Z} + a_{Z} [{\rm Mg/H}] \]
\[p = a_{0,tZ} + a_{t} \log(T_{\rm SSP}/10\,{\rm Gyr})  + a_{Z} [{\rm Mg/H}]  \ , \]
where $\sigma$ is velocity dispersion, $L_r$ is total r-band luminosity, $T_{\rm SSP}$ is spectroscopic age and [Mg/H] is spectroscopic Mg abundance, 
which is close to the total metallicity, Z/H. 
The coefficients $a$ are determined by {\it minimizing the scatter in colour}; this approach differs from some previous analyses (e.g. B11).

We quantify the fits primarily through the `coefficient of determination', $R^2$, defined as
the fraction of total variance in the colour which is accounted for by correlations with the predictor variable:
\[
R^2 = 1- \frac{ \sum_i (y_i - p_i )^2 }{  \sum_i (y_i - \bar{y} )^2 }  = 1- \frac{ \sigma^2_{\rm residual}} {\sigma^2_{\rm total}}
\]
Here, $y_i$ are the measured colours and $p_i$ are the colours predicted from a linear model, e.g.
in the case of  a fit to SSP-equivalent age. As indicated in the second expression, $1-R^2$ is the ratio of the residual variance around the fit to the 
total variance in the colour.

We also use a generalisation of the above to account for measurement errors: 
\[
T^2 = R^2 + \frac{ \sum_i e_{y,i}^2 + e_{p,i}^2 } {  \sum_i (y_i - \bar{y} )^2 }  = 
1- \left(\frac{ \sigma^2_{\rm residual}} {\sigma^2_{\rm total}} - \frac{\sigma^2_{\rm errors}} {\sigma^2_{\rm total}}\right)
\]
where $e_{y,i}$ are the errors in measured colour, and $e_{p,i}$ are the errors in the predicted colour, e.g. $e_p = a_t e_{\log t_{\rm SSP}}$ in the case of
the fit to age. 
$T^2$ thus represents the fraction of the total variance (in observed colour) which is attributable to a combination of 
(i) the correlations with the predictor variables and (ii) the measurement errors, both in colour and in the predictors.

\subsection{Observed colours versus  luminosity and velocity dispersion}

\begin{figure*}
\includegraphics[angle=0,width=180mm]{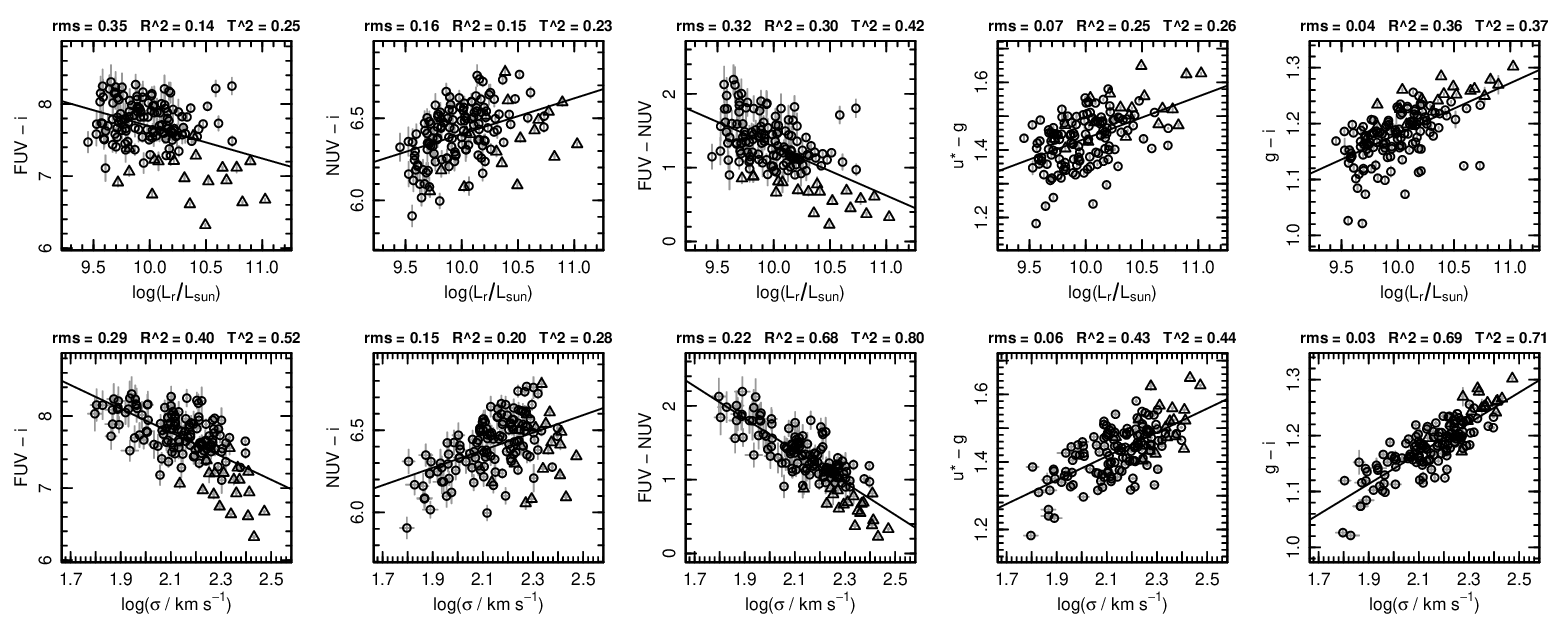}
\caption{UV and optical colours as a function of optical luminosity $L_{r}$ (upper panels) and velocity dispersion $\sigma$ (lower panels).
The solid line in each panel is the linear fit as recorder in Table~\ref{tab:fitcoeffs}. The header line indicates the rms scatter about the fit, 
the coefficient of determination $R^{2}$ (the fraction of total variance in the colour that is accounted for by the correlation), and the
$T^{2}$ parameter which is equivalent to $R^{2}$ but includes the effect of measurement errors. Galaxies marked with triangular symbols
have $FUV-NUV<0.9$, indicating that they are classical UV-upturn galaxies. 
}
\label{fig:cols_vs_mass}
\end{figure*}

\begin{figure*}
\includegraphics[angle=0,width=180mm]{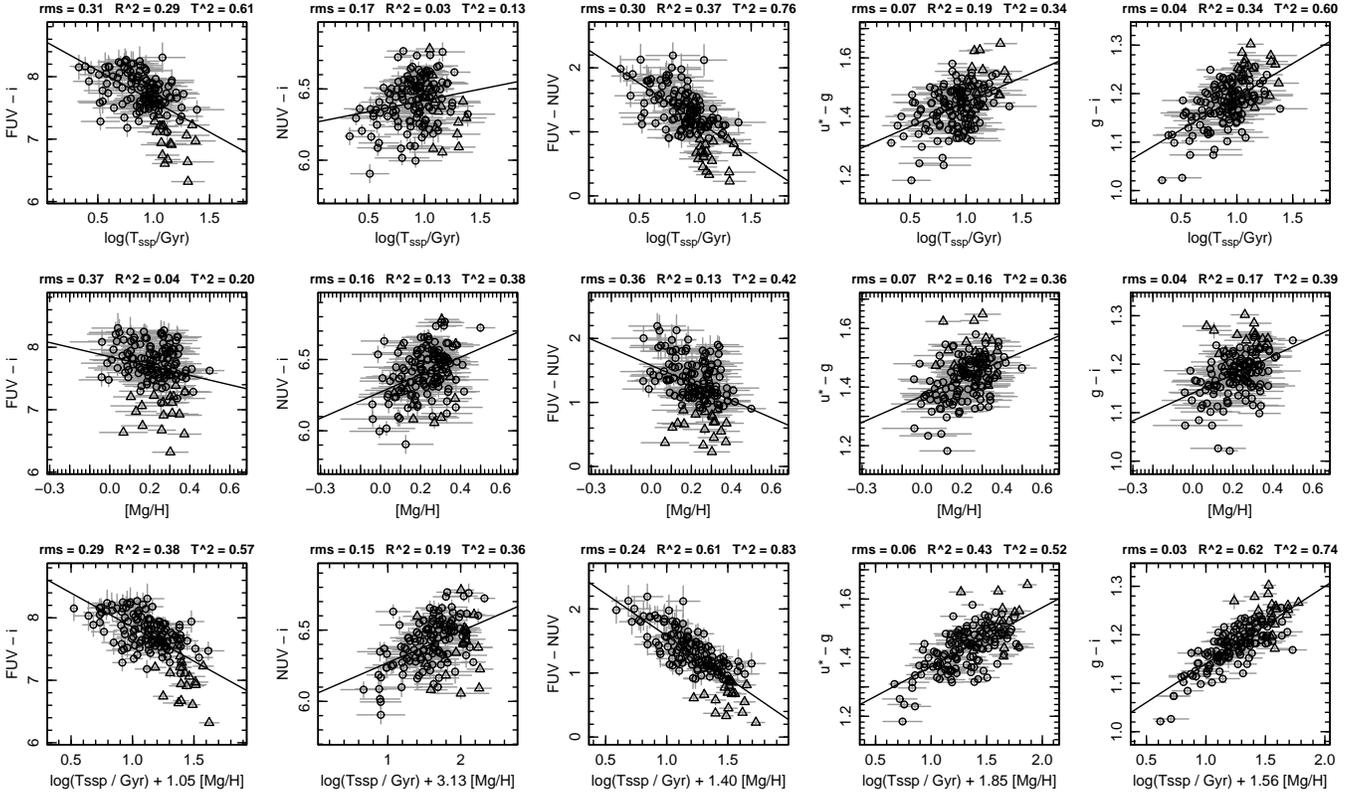}
\caption{UV and optical colours as a function of age ($T_{\rm \,SSP}$) and metallicity (Mg/H). 
Annotations are as in Figure~\ref{fig:cols_vs_mass}.
In the third row, the colours are plotted against the optimal linear combination of age and metallicity, as determined
from the simultaneous fits.
}
\label{fig:cols_vs_pops}
\end{figure*}

\begin{figure*}
\includegraphics[angle=0,width=170mm]{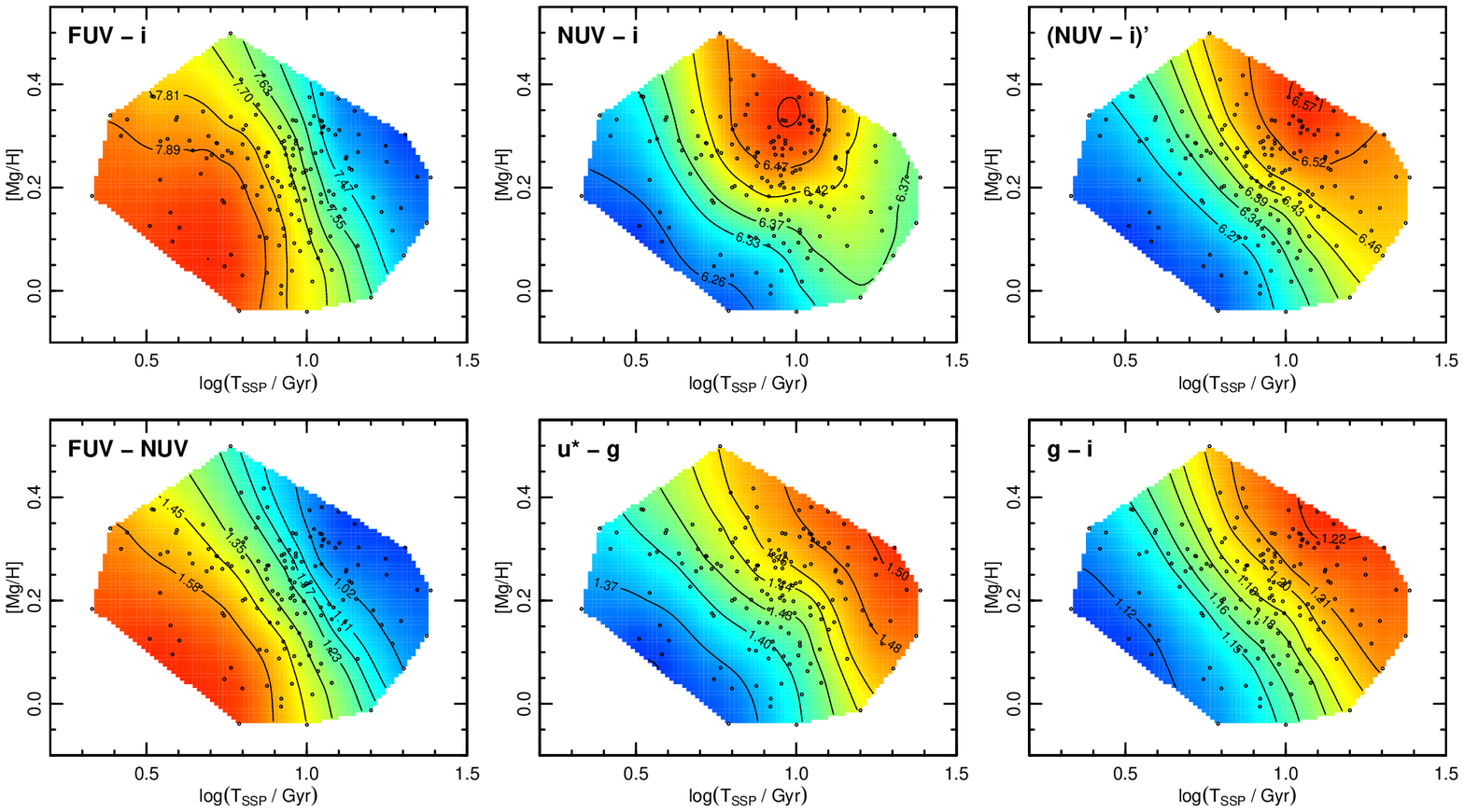}
\caption{
Maps of the average galaxy colour as a function of age and metallicity, created with a Kriging model (e.g. Cressie 1993). 
Black points indicate the galaxies used to generate the maps. The contours are located at deciles of the colour distribution.
The upper-right shows the results for $NUV-i$ after correcting for UV-upturn `leakage' in the NUV band (see Section~\ref{sec:leakage}).
}
\label{fig:fuvnuvkrig}
\end{figure*}

We begin by briefly considering the colour--luminosity and colour--$\sigma$ relations, as shown in Figure~\ref{fig:cols_vs_mass}.
The slopes and other parameters of the fits are reported in Table~\ref{tab:fitcoeffs}.

The colour--luminosity relations (upper panels of Figure~\ref{fig:cols_vs_mass}) show the familiar results that $FUV-i$ and $FUV-NUV$ become
{\it bluer} with increasing luminosity, contrary to the behaviour of colours at longer wavelengths. 
Galaxies with $FUV-NUV<0.9$ have a `classical' UV upturn, in the sense of having rising spectra ($f_\lambda$-vs-$\lambda$) in the UV regime. 
Such colours are seen mainly at the bright end of the luminosity distribution. 
Our approach is to treat the galaxy population continuously, rather 
than assign any particular significance to this value.
For reference however, we distinguish the 14 objects with $FUV-NUV<0.9$ using different symbols 
in this and subsequent figures\footnote{We note that Yi et al. (2011) found no bright Coma ellipticals satisfying $FUV-NUV<0.9$. 
The apparent discrepancy is probably due to the $k$-corrections applied in that work but not here.}.

The colour--luminosity relations show that the UV colours exhibit much greater scatter (0.2--0.3\,mag) than 
the optical colours ($\sim$0.05\,mag). 
In part, this difference in the absolute value of the scatter is trivially accounted for by the much larger total colour range spanned by the sample.
The quantities $R^2$ and $T^2$ scale this out by comparing the variance around the fit to the total variance in colour. The $R^2$ values
indicate that the luminosity dependence accounts for only $\sim$15 per cent of the total variance in  the `UV-to-optical' colours $FUV-i$ and $NUV-i$. 
Luminosity correlations account for a larger fraction of variance in the `optical-to-optical' colours $u^*-g$ (25 per cent) and $g-i$ (36 per cent).
Perhaps surprisingly, the luminosity trend also accounts for 30 per cent of the scatter in the `UV-to-UV' colour, $FUV-NUV$. 
We will find throughout this paper that $FUV-NUV$ follows tighter correlations than the UV-to-optical colours. 
The $T^2$ values show that measurement errors contribute an additional $\sim$10 per cent of the total variance in the UV colours, but make negligible
impact on the optical scatter. 

As noted in previous work (e.g. Marino et al. 2011; Carter et al. 2011), the trends of UV colours with velocity dispersion are tighter than those with luminosity.
The scalings with velocity dispersion, combined with the measurement errors,  account for around 70 per cent of the variance in $g-i$ and $FUV-NUV$, 
but only 30--50 per cent of the variance in $FUV-i$, $NUV-i$ and $u^*-g$. 
Fitting simultaneously to $\sigma$ and $L_r$ (not given in the table or figures) indicates there is no significant dependence on luminosity at fixed 
velocity dispersion, consistent with results in the optical (Bernardi et al. 2005; Graves et al. 2009). 

To summarize, all colours are strongly correlated (at the $\ga$5$\sigma$ level) with optical luminosity and velocity dispersion. 
These correlations account for a much of the observed variance in the $g-i$ and $FUV-NUV$ colours, but significant excess scatter remains
in $u^*-g$ and in the UV-to-optical colours.

\subsection{Observed colours versus age and metallicity}\label{sec:trends}

\begin{table*}
\caption{Parameters of linear fits to the observed colours. 
For each colour, the six lines give the results for 
(1) a null model fitting only a mean value; (2) a fit against luminosity; (3) a fit against velocity dispersion; 
(4) a fit against spectroscopic age; (5) a fit against spectroscopic metallicity; and (6) a simultaneous fit to age and metallicity. 
The coefficients $a$ refer to a general model of the form 
$p = a_{0} + a_L \log(L_r/(10^{10}L_\odot)) + a_\sigma \log(\sigma/100\,{\rm km\,s^{-1}}) + a_t \log(t_{\rm SSP}/10\,{\rm Gyr}) + a_Z [\rm {Mg/H}]$, 
although only subsets of these terms are used in each fit.
For each line, the quantity $R^2$ is the ratio of the total variance explained by the fitted model.
$T^{2}$ is the fraction of variance explained by the fitted model {\it and} 
the nominal measurement errors in both the colours and the predictor variables. 
} 
\begin{tabular}{llccccccc}
\hline
Colour & $a_0$ & $a_L$ & $a_\sigma$ & $a_t$ & $a_Z$ & rms & $R^2$ & $T^2$ \\
\hline

~\\
 $FUV - i$ & $7.683\pm0.002$ & --- & --- & --- & --- & 0.373 & --- & --- \\
          & $7.691\pm0.028$ & ${-0.447\pm0.090}$ & --- & --- & --- & 0.346 & 0.14 & 0.25  \\
          & $7.937\pm0.035$ & --- & ${-1.643\pm0.163}$ & --- & --- & 0.287 & 0.41 & 0.52  \\
          & $7.605\pm0.028$ & --- & --- & ${-0.987\pm0.126}$ & --- & 0.314 & 0.29 & 0.61  \\
          & $7.850\pm0.071$ & --- & --- & --- & ${-0.752\pm0.290}$ & 0.365 & 0.04 & 0.21  \\
          & $7.851\pm0.057$ & --- & --- & ${-1.092\pm0.119}$ & ${-1.144\pm0.236}$ & 0.291 & 0.39 & 0.58  \\
~\\
 $NUV - i$ & $6.407\pm0.024$ & --- & --- & --- & --- & 0.171 & --- & --- \\
          & $6.404\pm0.013$ & ${+0.217\pm0.041}$ & --- & --- & --- & 0.157 & 0.16 & 0.24  \\
          & $6.325\pm0.018$ & --- & ${+0.534\pm0.087}$ & --- & --- & 0.153 & 0.20 & 0.28  \\
          & $6.420\pm0.015$ & --- & --- & ${+0.158\pm0.067}$ & --- & 0.168 & 0.04 & 0.14  \\
          & $6.272\pm0.031$ & --- & --- & --- & ${+0.612\pm0.126}$ & 0.159 & 0.14 & 0.38  \\
          & $6.271\pm0.030$ & --- & --- & ${+0.221\pm0.062}$ & ${+0.691\pm0.124}$ & 0.153 & 0.20 & 0.37  \\
~\\
$FUV - NUV$ & $1.275\pm0.012$ & --- & --- & --- & --- & 0.382 & --- & --- \\
          & $1.287\pm0.026$ & ${-0.664\pm0.083}$ & --- & --- & --- & 0.320 & 0.30 & 0.42  \\
          & $1.613\pm0.026$ & --- & ${-2.177\pm0.123}$ & --- & --- & 0.216 & 0.68 & 0.81  \\
          & $1.185\pm0.026$ & --- & --- & ${-1.145\pm0.121}$ & --- & 0.302 & 0.38 & 0.76  \\
          & $1.578\pm0.069$ & --- & --- & --- & ${-1.364\pm0.282}$ & 0.355 & 0.14 & 0.42  \\
          & $1.580\pm0.046$ & --- & --- & ${-1.312\pm0.097}$ & ${-1.835\pm0.192}$ & 0.237 & 0.62 & 0.83  \\
~\\
 $u^* - g$ & $1.436\pm0.019$ & --- & --- & --- & --- & 0.077 & --- & --- \\
          & $1.434\pm0.005$ & ${+0.124\pm0.017}$ & --- & --- & --- & 0.067 & 0.26 & 0.26  \\
          & $1.382\pm0.007$ & --- & ${+0.353\pm0.033}$ & --- & --- & 0.058 & 0.44 & 0.45  \\
          & $1.449\pm0.006$ & --- & --- & ${+0.168\pm0.028}$ & --- & 0.069 & 0.20 & 0.34  \\
          & $1.370\pm0.014$ & --- & --- & --- & ${+0.300\pm0.056}$ & 0.071 & 0.16 & 0.36  \\
          & $1.369\pm0.011$ & --- & --- & ${+0.202\pm0.024}$ & ${+0.373\pm0.047}$ & 0.058 & 0.44 & 0.53  \\
~\\
   $g - i$ & $1.183\pm0.005$ & --- & --- & --- & --- & 0.048 & --- & --- \\
          & $1.181\pm0.003$ & ${+0.091\pm0.010}$ & --- & --- & --- & 0.038 & 0.36 & 0.38  \\
          & $1.141\pm0.003$ & --- & ${+0.273\pm0.015}$ & --- & --- & 0.026 & 0.69 & 0.71  \\
          & $1.194\pm0.003$ & --- & --- & ${+0.137\pm0.015}$ & --- & 0.038 & 0.35 & 0.61  \\
          & $1.140\pm0.008$ & --- & --- & --- & ${+0.191\pm0.034}$ & 0.043 & 0.17 & 0.40  \\
          & $1.140\pm0.006$ & --- & --- & ${+0.160\pm0.012}$ & ${+0.249\pm0.023}$ & 0.029 & 0.63 & 0.74  \\

\hline
\end{tabular}
\label{tab:fitcoeffs}
\end{table*}

\begin{table*}
\caption{Comparision of fits using age and Mg/H to those using age and Fe/H. An intercept term has also been fitted but is not reported here.}
\begin{tabular}{lccccccccc}
\hline
& \multicolumn{4}{c}{Fitting age and Mg/H} & 
& \multicolumn{4}{c}{Fitting age and Fe/H} \\
Colour \ \ \ \ \ \ \ \ \ \ \ \ \ \ \ \ \ \ \ \ \ 
& $a_t$ & $a_Z$ & rms & $R^2$ & \ \ \ \ \ 
& $a_t$ & $a_Z$ & rms & $R^2$ \\
\hline

$FUV - i$            & $-1.092\pm0.119$ & $-1.144\pm0.236$ & 0.291 & 0.38 && $-1.092\pm0.159$ & $-0.267\pm0.250$ & 0.312 & 0.29 \\
$NUV - i$            & $+0.221\pm0.062$ & $+0.691\pm0.124$ & 0.153 & 0.19 && $+0.488\pm0.073$ & $+0.845\pm0.115$ & 0.144 & 0.28 \\
$FUV - NUV$          & $-1.312\pm0.097$ & $-1.835\pm0.192$ & 0.237 & 0.61 && $-1.580\pm0.142$ & $-1.112\pm0.224$ & 0.279 & 0.46 \\
$u^* - g$            & $+0.202\pm0.024$ & $+0.373\pm0.047$ & 0.058 & 0.43 && $+0.288\pm0.031$ & $+0.307\pm0.049$ & 0.062 & 0.36 \\
$g - i$              & $+0.160\pm0.012$ & $+0.249\pm0.023$ & 0.029 & 0.62 && $+0.225\pm0.016$ & $+0.225\pm0.025$ & 0.031 & 0.58 \\

\hline
\end{tabular}
\label{tab:metfits}
\end{table*}

We assume that the trends of colour with luminosity and velocity dispersion result from underlying variation in the {\it average} 
stellar populations in galaxies as a function of these quantities. In this section we use the spectroscopic ages and metallicities 
to explore the degree to which such correlations can explain the observed colour variation.

In the absence of photometric errors (and of internal reddening or non-stellar flux sources such as active nuclei), the colours should be
completely determined from the stellar populations present. In practice however, we do not have full information about these populations.
The spectroscopic data provide only a single characteristic age and metallicity, and estimates for abundance ratios of a few elements. 
At best, these are only moments of the true distribution in these parameters (e.g. $T_{\rm SSP}$ is weighted to the age of the most recent star-formation 
episodes). An important caveat, in the context of this paper, is that if the UV colours are driven by sub-populations far from the average 
population, such as the very oldest or youngest stars, or the low- or high-metallicity `tail' of the metallicity distribution, then this may not be reflected cleanly 
in correlations with average SSP-equivalent properties.
However, spectroscopic age and metallicity are still more likely to reflect meaningful information about the galaxies than using individual line index strengths, 
which always depend on combinations of these parameters, as well as on individual element abundances.
At worst, the SSP-equivalent parameters may be influenced by effects that are not adequately taken into account in the analysis used to derive them. 
For example, the measured `age' may partly depend on the incidence of blue horizontal-branch stars 
(e.g. Lee, Yoon \& Lee 2000; Schiavon et al. 2004; Percival \& Salaris 2011), as well
as the main-sequence turnoff mass. Even in this case, however, the optical age and metallicity 
provide a useful compression of the spectroscopic constraints on the stellar content that are available from the optical light.

To test for correlations between the colours and the stellar populations which dominate the spectra,
we fit colour versus age and metallicity (Mg/H), first individually, and then in linear combination. 
We use Mg/H as the primary metallicity tracer, rather than Fe/H, 
since the abundance of $\alpha$ elements better reflects the total metallicity Z/H (due to the large contribution of oxygen to the total number fraction in metals).
The results are shown in Figure~\ref{fig:cols_vs_pops}, and the fit results summarised in Table~\ref{tab:fitcoeffs}. 

The first row of panels in Figure~\ref{fig:cols_vs_pops} show that $u^*-g$, $g-i$ and $NUV-i$ all become redder with increasing age, while $FUV-NUV$
and $FUV-i$ become bluer with increasing age. The coefficient of age is significant at the $>$6$\sigma$ level for all colours except $NUV-i$, where the correlation
is only marginally significant (2.4$\sigma$). 
In $FUV-i$, we note an apparent steepening of the age dependence among the oldest galaxies. For $T_{\rm ssp}${}$>$9\,Gyr (the median age of the sample galaxies),
the slope is $-1.50\pm0.35$\,mag per decade in 
age, while for $T_{\rm ssp}${}$<$9\,Gyr the slope is shallower and barely significant, at $-0.39\pm0.21$\,mag per decade. 
According to $R^2$ (i.e. without accounting for measurement errors), the linear
age trends make 20--40 per cent contributions to the total variance in the colours (except $NUV-i$).
However, the error contribution is substantial, because age has fairly large uncertainty, 
compared to its range. Using $T^2$ to incorporate the errors, we account for over 60 per cent of the variance in $FUV-i$, $FUV-NUV$  and $g-i$, but less than 
half of the variance in $u^*-g$ and $NUV-i$.

The second row of panels in Figure~\ref{fig:cols_vs_pops}  shows the equivalent correlations with metallicity. These are
in the same sense as the age trends: colours which redden with increasing age also redden with increasing metallicity.
The coefficients are significant at the $\sim$5$\sigma$ level for all colours except $FUV-i$ (2.6$\sigma$).
Again, measurement errors contribute substantially to the scatter, since errors on Mg/H individually are large. 
Based on $T^2$, the metallicity trends and measurement errors account for about 40\ per cent of the variance in most colours, though only 20 per cent in $FUV-i$.

Above, we have modelled the colours as a function {\it either} of age or of metallicity, but  these two parameters are not truly independent. 
In particular, at fixed velocity dispersion or luminosity, 
older galaxies tend to be more metal poor (e.g. Trager et al. 2000; Smith et al. 2009). 
This anti-correlation appears to have an intrinsic component, but there is also a contribution from
error covariance, due to the `tilt' of model grids relative to the line-strength indices. Regardless of its source, 
the mutual correlation of age and metallicity makes it difficult to determine, from single-parameter fits, 
which parameters are responsible for driving the observed colours. Moreover, population synthesis models
show that optical colours, at least, are sensitive to both age and metallicity in similar degree. 
Hence to disentangle the effects of age and metallicity, we fit two-parameter models of the form 
$p =  a_0 + a_t \log(t_{\rm SSP}/10\,{\rm Gyr}) + a_Z [\rm {Mg/H}]$, 
to identify an improved predictor of each colour that combines the age and metallicity effects to minimize the scatter.

In the lower panels of Figure~\ref{fig:cols_vs_pops}, we express this combination as $\log(t_{\rm SSP}/10\,{\rm Gyr}) + S [\rm {Mg/H}]$,
where the ratio $S=a_Z/a_t$ differs for each colour. For the optical colours $u^*-g$ and $g-i$, we recover $S=1.6-1.8$, 
in agreement with $S=1.6$ expected from the `three-halves rule'  of Worthey (1994). Note, however, that Worthey's results were derived from
the colours {\it predicted} by population synthesis models with known age and metallicity. Here, we have instead derived this ratio on the basis
of {\it data} for colour,  age and metallicity\footnote{The fact that our SSP parameters were derived via comparison to synthesis models does not 
make this a circular argument, since only spectroscopic indices were used to fit the ages, with no information about the broadband colours.}.
For the UV colours, we find that $NUV-i$ is  more sensitive to metallicity than to age, compared to the optical colours ($S\approx3.1$), while $FUV-i$ is 
somewhat more age sensitive ($S\approx1.0$). All of the colours studied show significant dependence on both age and metallicity in the simultaneous fits. 
In Table~\ref{tab:metfits} we compare the fit results obtained if Mg/H is replaced with Fe/H as the metallicity indicator.  
For most colours, there is little change to the fit results, although the scatter is generally slightly increased when using Fe/H. 
For $NUV-i$, however, we recover a slightly reduced scatter (marginally significant), and a steeper dependence on age, when using Fe/H. 
This is likely due to strong line blanketing in the NUV region, which is thought to be dominated by iron lines (Peterson, Dorman \& Rood 2001).

Figure~\ref{fig:fuvnuvkrig} presents these trends in an alternative manner, as maps of average galaxy colour in the age--metallicity plane. 
The slope of the contours in the maps reflects the relative sensitivity of each colour to age and metallicity. The map for $NUV-i$ appears to show
distorted contours indicating a more complex behaviour for this colour. We argue in Section~\ref{sec:leakage} that this is due to `contamination' 
of the NUV by the old hot stars responsible for the UV upturn. 

In most cases, the simultaneous fit to age and metallicity accounts for substantially more of the total variance in colour
than either parameter individually, and the residual scatter is comparable to that around the colour-$\sigma$ relations. 
The effect of measurement errors is smaller than for the age or metallicity taken separately, 
because the age errors are anti-correlated with the metallicity errors, and so tend to compensate in the predicted colours.

\subsection{Systematics}\label{sec:systematics}

In this section we consider possible systematic error sources, and test whether they can make significant contributions to the residual scatter 
around the fits.

\subsubsection{Calibration uncertainties}\label{sec:calerrs}

The aperture magnitude errors adopted for the analysis in Section~\ref{sec:trends} do not include contributions from systematic calibration uncertainties, because most galaxies in the sample 
are drawn from only a small number of observations (\galex\ tiles, SDSS drift-scan `runs', MegaCam pointings). Hence calibration errors 
are likely to affect large sections of the dataset coherently, rather than on a galaxy-by-galaxy basis. 
In this section, we assess the evidence for constant shifts 
between individual `observations', and their impact on the colour scatter, by allowing additional offset terms in the fits. 

For the $g-i$ colour, where SDSS is the only data source, $\sim$90\,per cent of the galaxies were observed in only two drift-scan runs (numbers 5115 and 5087).
Including a term representing offsets between runs, we find that colours measured from run 5087 are
on average bluer  by $0.009\pm0.005$\,mag than those measured from run 5115, at similar age and Mg/H. This is consistent with the claimed calibration errors of 0.01\,mag 
in $g$ and $i$ (Padmanabhan et al. 2008). Although this term is marginally significant, its inclusion
does not reduce the residual scatter around the $g-i$ fit, which remains 0.029\,mag. 
A similar test allowing offsets at the level of SDSS camera columns {\it within} each run does not
reveal any further significant systematic effects at above the $\sim$0.01\,mag level.

To test for calibration shifts in the UV versus optical colours, we fit a similar offset term for each \galex\ tile. Here $\sim$80\,per cent of the galaxies are drawn from 
two tiles (GI5\_025001\_COMA  and  GI2\_046001\_COMA3). The offset terms are not significant: GI2\_046001\_COMA3 yields colours on average bluer by $0.04\pm0.06$\,mag
in $FUV-i$ and redder by $0.01\pm0.03$\,mag in $NUV-i$, for galaxies of similar age and metallicity. 
Again, these limits are consistent with the nominal \galex\ calibration uncertainties of 0.05\,mag (FUV) and 0.03\,mag (NUV) 
quoted by Morrissey et al. (2007). Allowing offset terms does not reduce the residual scatter. 

Applying a similar approach to the $u^*-g$ colours reveals that significant systematic errors are present in the MegaCam photometry. 
Allowing offset terms associated with each of the nine MegaCam pointings, we find that the pointing covering the south-west of Coma 
(contributing 12\,per cent of the sample galaxies) yields colours $0.09\pm0.01$ bluer in $u^*-g$ than the central pointing (57\,per cent of the sample),
for galaxies of similar age and metallicity. That this is due to calibration errors in  MegaCam, rather than SDSS, is confirmed by null results
obtained fitting SDSS run offsets (as for $g-i$), either separately, or in combination with MegaCam pointing offsets. 
The possibility that the offset in the south-west MegaCam pointing is `astrophysical' rather than systematic can be rejected given that no corresponding shifts
are seen for south-west region galaxies in residuals for the other colours ($g-i$, $NUV-i$, etc). Allowing for MegaCam offsets reduces the 
residual scatter in $u^*-g$ from 0.058\,mag to 0.049\,mag and the explained fraction of variance increases to $R^2=0.63$ (from 0.44). 

\subsubsection{K-corrections}\label{sec:kcorr}

We did not apply $k$-corrections to the measured colours, due to the uncertainty in the SEDs, especially in the UV. 
Instead, for each colour we have tested for contributions due to band-shifting, by explicitly including an extra model term proportional to 
redshift. 

A significant effect ($\sim$5$\sigma$ level) is recovered only for the $u^*-g$ colour.  
The redshift dependence of $u^*-g$ is not unexpected, since the red cut-off of the $u^*$ filter is close to the 4000\,\AA\ break shifted to the mean
recession velocity of Coma. 
Allowing for this dependence increases the explained  fraction of variance in $u^*-g$ by a further 6 per cent. 
Including both the redshift term and the MegaCam systematics discussed above, the explained fraction of variance reaches $R^2=0.69$, 
even without allowing for random errors. 

While marginal (2--3$\sigma$) $k$-correction terms are also
detected in the residuals for $g-i$ and $FUV-NUV$, they contribute negligibly to the variance (1--3 per cent) in these colours.

\subsubsection{Aperture mismatch}\label{sec:mismatch}

Our fits compare colours derived in 10\,arcsec diameter photometric apertures against age and metallicity estimates derived from spectroscopy in much smaller
apertures (3\,arcsec diameter in most cases). An obvious concern is whether mismatch between these apertures contributes to the scatter. 
This is a particular concern in the UV, which would be sensitive to any ongoing star-formation occurring beyond the spectrograph fibre (e.g. Salim \& Rich 2010).
As for photometric calibration shifts and $k$-corrections, we can constrain these effects by extending our fit  with additional terms relating to aperture mismatch.  

Since the GALEX resolution does not allow use of smaller apertures, we use the SDSS photometry to define a colour shift 
$\Delta_{gi} = (g-i)_{\rm 10^{\prime\prime}} - (g-i)_{\rm 3^{\prime\prime}}$. On average, $\Delta_{gi}$ is negative, indicating that the sample galaxies are typically 
bluer in $g-i$ within the photometric aperture than within the spectroscopic aperture. This is the sense expected from metallicity gradients in purely passive galaxies.

We constrain the effect of aperture mismatch in our fits by including an additional term proportional to $\Delta_{gi}$. For $FUV-i$, the recovered coefficient
is not statistically significant, at $-3.1\pm2.1$. The negative coefficient indicates that galaxies which are bluer in the photometric aperture than in the fibre in
$g-i$ are globally {\it redder} in $FUV-i$ than average for galaxies of similar age and metallicity. This is {\it opposite} to the sense expected if the $FUV-i$ colour scatter 
were dominated by star formation beyond the spectroscopic fibre. Including the  $\Delta_{gi}$ term in fit has negligible impact on the 
residual scatter which remains 0.29\,mag. 
For $NUV-i$, the recovered coefficient is positive but not significant ($+1.4\pm1.1$), and its inclusion does not reduce the scatter around the fit, which remains 0.15\,mag. 
For the optical colours, there is a small positive correlation with $\Delta_{gi}$, with no associated reduction in scatter, similar to the case for $NUV-i$. 
Repeating these tests using colour shift in $u^*-g$ (i.e. $\Delta_{ug}$ instead of $\Delta_{gi}$) as an indicator of aperture mismatch
yields no further evidence for star formation beyond the fibre aperture, nor any non-negligible reduction in scatter around the fits. 

We conclude that unidentified star formation beyond the spectrograph fibre contributes negligibly to the observed colour scatter at fixed age and metallicity.  

\subsubsection{Internal extinction}

Although we cannot exclude the possibility that variations in internal extinction 
impose some additional scatter in the relationships, it is clear that this does not dominate the residuals. For example, the residual scatter
of 0.15\,mag in $NUV-i$, if attributed wholly to extinction, would imply a minimum rms of 0.038\,mag in $g-i$, which is far in excess of the 
0.029\,mag observed residual scatter\footnote{This calculation assumes the Cardelli, Clayton \& Mathis (1989)
$R_{V}=3.1$ extinction law.}. Moreover, the residuals in $FUV-i$ are anti-correlated with those in $g-i$: galaxies that are redder than average (for their
age and metallicity) in the optical are unexpectedly blue in $FUV-i$. Such a relationship cannot be generated only by internal extinction variations. 
Finally, the distribution of the UV versus optical colour residuals is skewed, with a tail of galaxies extending towards bluer colours. This could be caused only by 
these galaxies having unusually {\it low} dust content relative to the average, while the dispersion of internal extinction about its average value remains small. 
We consider this unlikely.

\subsubsection{Summary}

In summary, having reviewed the likely sources of systematic errors, we find evidence for substantial effects only in the $u^*-g$ colours, due to photometric
offsets and band-shifting with redshift. Including these systematic errors, the correlations with spectroscopic age and metallicity 
account for 70--80 per cent of the  variance in the optical colours and in $FUV-NUV$, but still only 40--60 per cent of the variance in $FUV-i$ and $NUV-i$. 
 
\subsection{Residual correlations with abundance ratios}

\begin{figure*}
\includegraphics[angle=0,width=180mm]{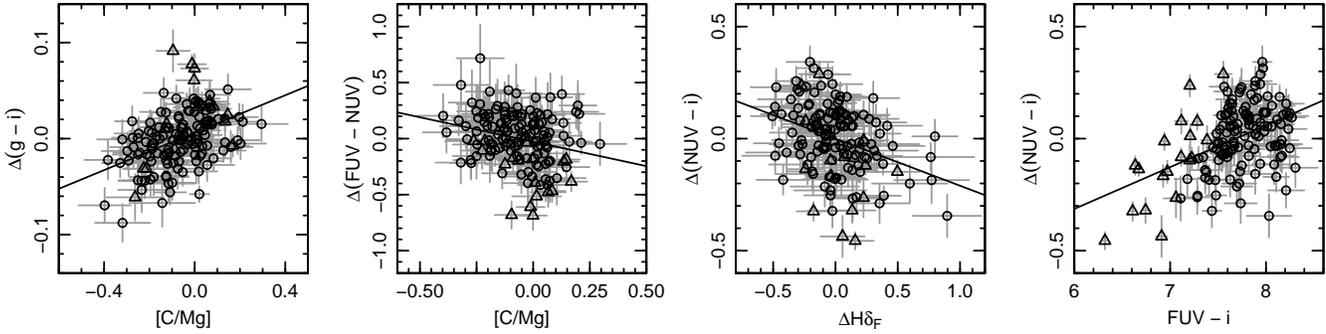}
\caption{Residual correlations discussed in the text. 
The residuals $\Delta(NUV-i)$, etc, are as measured from the fits using age and metallicity simultaneously
to predict the colours. The vertical error-bars include the errors in the predictor variables, corrected for covariance as necessary.}
\label{fig:resids}
\end{figure*}

We have shown that, using the measured ages and metallicities, we are able to predict optical/UV colours of passive galaxies 
with the same level of precision as using the velocity dispersion. This suggests that the trends of {\it mean} colour with $\sigma$ 
are adequately reproduced by the trends of {\it mean} spectroscopic age and metallicity with $\sigma$. 
However, for $FUV-i$ and $NUV-i$, where much of the scatter in colour was unexplained by correlation with velocity dispersion,
this scatter was not reduced by using models based on the spectroscopic constraints.

Since element abundances in red-sequence galaxies cannot be adequately described by a single overall metallicity (e.g. Worthey 1998), the abundance ratios 
of individual elements are obvious candidates for additional parameters which might contribute to driving the colours. In the case of the UV upturn, 
in particular, the correlation of $FUV$-vs-optical colours with magnesium line strength indices, and relative weakness of the correlation with iron-dominated indices,
has led many authors to speculate that the UV upturn depends primarily on $\alpha$-element abundance (O'Connell et al. 1999; Carter et al. 2011; B11). 

We have tested for correlations with the element abundances via linear fits of the colour residuals (from the age--metallicity models) against 
the ratios Fe/Mg, Ca/Mg, C/Mg and N/Mg. (We express all the ratios relative to Mg, rather than the more usual Fe, 
since we have adopted Mg/H as our overall metallicity indicator.) 
We fit simultaneously to all the ratios, to disentangle more cleanly the effects of individual elements.
The results are summarized in Table~\ref{tab:resids_vs_xfe}.

\begin{table*}
\caption{
Residual correlations with abundance ratios. 
Note that because these are fits to the {\it residuals} from the age--metallicity fits, the coefficients describe the effects of the given abundance ratio at
fixed Mg/H. Coefficients that are significant at the $>3\sigma$ level are highlighted with bold type. 
The final columns show the small reduction in total scatter when the abundance ratio trends are included in the fit. 
  } 
\begin{tabular}{lcccccc}
\hline
colour residual & [Fe/Mg] & [C/Mg] & [N/Mg] & [Ca/Mg] & original rms & new rms \\ 
\hline 
$\Delta$($FUV - i$)	&	 ${+0.67\pm0.26}$	&	 ${-0.33\pm0.20}$	&	 ${-0.12\pm0.18}$	&	 ${-0.15\pm0.30}$	&	 0.291	&	 0.284	\\
$\Delta$($NUV - i$)	&	 ${+0.22\pm0.13}$	&	 ${\bf{}+0.30\pm0.10}$	&	 ${+0.23\pm0.09}$	&	 ${+0.00\pm0.15}$	&	 0.153	&	 0.141	\\
$\Delta$($FUV - NUV$)	&	 ${+0.46\pm0.21}$	&	 ${\bf{}-0.63\pm0.16}$	&	 ${-0.36\pm0.14}$	&	 ${-0.15\pm0.23}$	&	 0.237	&	 0.222	\\
$\Delta$($u^* - g$)	&	 ${+0.02\pm0.05}$	&	 ${+0.10\pm0.04}$	&	 ${+0.07\pm0.04}$	&	 ${-0.02\pm0.06}$	&	 0.058	&	 0.056	\\
$\Delta$($g - i$)	&	 ${+0.00\pm0.02}$	&	 ${\bf{}+0.10\pm0.02}$	&	 ${+0.04\pm0.02}$	&	 ${+0.01\pm0.03}$	&	 0.029	&	 0.025	\\

\hline
\end{tabular}
\label{tab:resids_vs_xfe}
\end{table*}

We find only marginal correlations of colours with abundance ratios in most cases. 
For the Fe/Mg ratio, we find that $FUV-i$ colour reddens by 0.07\,mag, and $FUV-NUV$ reddens by 0.05\,mag, 
for a 0.1\,dex increase in Fe/Mg at fixed age and Mg/H (i.e. bluer UV-vs-optical colours for Mg-enhanced populations).
However, these correlations are only $\sim$2.5$\sigma$ effects.

Among the other residual trends with abundance ratios, the only significant correlations are for C/Mg: 
a 0.1\,dex enhancement in this ratio (at fixed age and Mg/H) is associated with a 0.01\,mag reddening of
$g-i$, a 0.03\,mag redenning in $NUV-i$ and a 0.06\,mag shift bluewards in $FUV-NUV$ (all at 3--4$\sigma$). 
These residual trends are shown in the first two panels of Figure~\ref{fig:resids}.

To test the robustness of these results, we have also used an alternative approach in which individual abundance ratio terms
are incorporated into the fit for each colour as extra parameters, in addition to age and metallicity. 
Including only Fe/Mg as a third parameter, 
we find significant trends for $FUV-i$ ($+0.80\pm0.30$), $NUV-i$ ($+0.69\pm0.16$) and $g-i$ ($+0.11\pm0.03$). 
Repeating this test with C/Mg as a third parameter instead of Fe/Mg, significant trends are found for 
 $NUV-i$ ($+0.33\pm0.09$), $FUV-NUV$ ($-0.45\pm0.15$) and $g-i$ ($+0.10\pm0.02$). 
 For $u^*-g$, a significant trend with C/Mg ($+0.16\pm0.03$) is recovered after accounting for the systematic effects noted in Section~\ref{sec:systematics}.
These results are qualitatively similar to those obtained from the simultaneous fit to the residuals, although individual coefficients
are sensitive to the fitting treatment adopted. 

To summarize, decoupling the effects of multiple abundance ratio parameters, in addition to the age and metallicity trends, 
remains difficult. The strongest abundance ratio effects in the UV seem to be related to Fe/Mg and C/Mg, with little dependence
on Ca/Mg or N/Mg.  
The abundance ratio effects make modest contributions to the total observed scatter, increasing $R^2$ by 8 per cent for $NUV-i$ and $g-i$, 
and by 4--6 per cent in $FUV-i$, $FUV-NUV$ and $u^*-g$.

\section{Discussion}\label{sec:discuss}

\subsection{The index-vs-colour relations reconsidered}\label{sec:burstest}

As reviewed in the Introduction, early work revealed a strong correlation between the FUV-vs-optical colour and the Mg$_2$ line index
(Burstein et al. 1988), from which it was inferred that the FUV output of passive galaxies 
depends mainly on metallicity. Equivalent conclusions were drawn by B11, using much improved
data, from the strong correlation of the Mg$b$ index against $FUV-V$ colour. 
In this section we address the apparent disagreement between this view and 
the results of our analysis which instead favour a dominant age dependence of
the FUV-vs-optical colour, and only a weaker residual trend with metallicity. 
We base our argument on comparison with B11, but qualitatively the same case applies to 
Burstein et al. (1988) and other studies following a similar analysis approach. 

\begin{figure}
\includegraphics[angle=0,width=82mm]{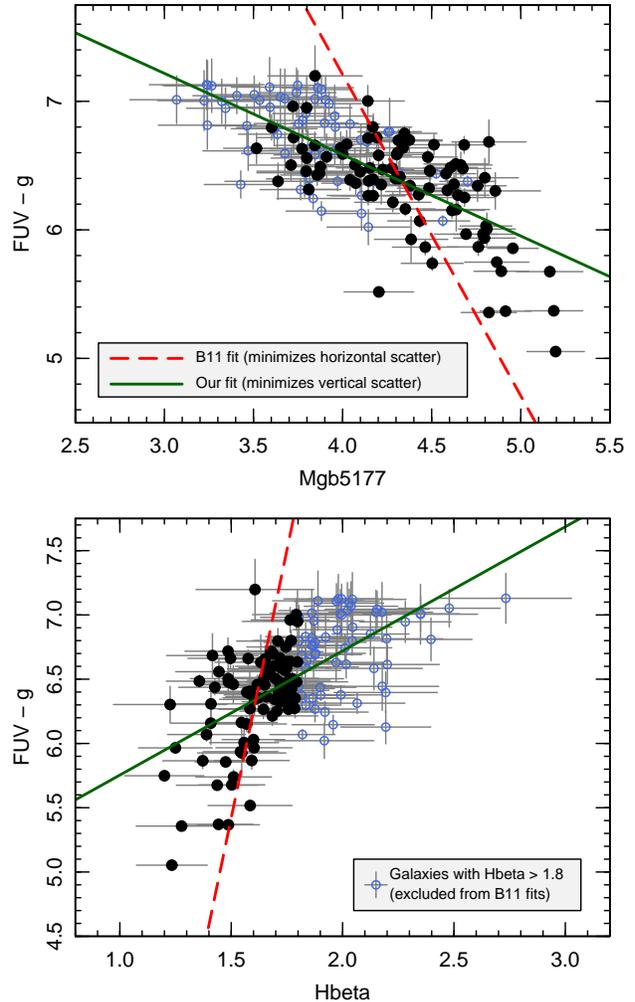}
\caption{The linestrength-vs-colour relations, for comparison to previous work, in particular Bureau et al. (2011, B11). 
The green line shows our fit to our data, minimizing in the colour direction, as throughout
our work. The red line shows the B11 fits (to their data), which minimize in the line-strength direction and exclude 
galaxies with H$\beta$$>$1.8\,\AA. 
}
\label{fig:burplot}
\end{figure}

In Figure~\ref{fig:burplot}, we show the correlations of $FUV-g$ (to match B11's $FUV-V$) with the Mg$b$ and H$\beta$ line strength indices. 
As before, we treat {\it colour} as the dependent variable, and fit the relations by minimizing scatter in the colour direction. 
For comparison we show also the fits quoted by B11. Their approach minimizes scatter in the {\it line-strength} direction and their fits are made only to 
galaxies with H$\beta<1.8$\,\AA. To account for the different colour definitions used, the B11 fits have been shifted to pass through the 
average position of our H$\beta<1.8$\,\AA\ galaxies, but the slopes are as quoted in their paper. 
As always for correlations with substantial scatter, the fits differ significantly according to the direction of minimization. 
Fitting our data in the same way as B11 fit theirs, we would recover similar slopes. Hence at the level of the data, our index-vs-colour correlations
are consistent with earlier work. 

The origin of the different conclusions we reach, with respect to B11, lies in three differences in our analysis: 
the choice of minimization direction, the treatment of the high H$\beta$ data-points and the interpretation of the recovered trends. 

Minimizing in the colour direction implies asking which parameters are the best {\it predictors} of colour. When considering age and metallicity as
predictors (as in Section~\ref{sec:trends}, there are strong grounds for assuming a causal relationship, since
these parameters at least in part determine the stellar content, and hence the
broadband colours. However, even individual line-strength indices are related to the properties of the optically-dominant  stellar population
in a fairly direct way. The UV colours by contrast have long appeared to exhibit large scatter with respect to all of the optical properties. 
Hence colour is the quantity with `unknown' behaviour, 
 
 and consequently we prefer to treat it as the dependent variable. 

The effect of excluding  strong-H$\beta$ (i.e. young and/or metal-poor) galaxies itself depends on the minimization direction. Minimizing residuals in colour, as 
in our approach, the derived trend is unchanged by such a cut, to first order, even when fitting H$\beta$ itself. By contrast, if we minimize residuals in 
index strength (as done by B11), the resulting fit is highly biased for H$\beta$, since all H$\beta$ values are included at blue colours, but only unrepresentatively  
low H$\beta$ values contribute at  redder colours. The bias is less strong for Mg$b$, since the selection is not in the fitted quantity, but still
present because Mg$b$ tends to be anti-correlated with H$\beta$. (Note that our fits may be slightly affected by a bias in the opposite sense: 
since our sample is limited by magnitude error in the FUV, the reddest galaxies may be preferentially excluded at the faint end.) 

Thus we can understand why the different choice of fit direction and exclusion of strong-H$\beta$ galaxies led B11 to a steeper slope
for the colour-vs-Mg$b$ relation (inverting the values in their table 2 gives $-2.5^{+0.5}_{-0.7}$\,mag\,\AA$^{-1}$ compared to our slope
of $-0.63\pm0.05$\,mag\,\AA$^{-1}$) and a dramatically steeper slope 
for the colour-vs-H$\beta$ relation ($+8.3^{+5.9}_{-2.5}\,$mag\,\AA$^{-1}$ compared to our $+0.97\pm0.10$\,mag\,\AA$^{-1}$).
It might be expected, then, that B11 would infer a much steeper age dependence for the FUV-vs-optical colour than we do. 
Instead, they imply that the trend results from the known dependence on H$\beta$ on metallicity, and hence that a correlation
between FUV and metallicity drives both the colour-Mg$b$ and colour-H$\beta$ relationships. 
B11 do acknowledge that  the high-H$\beta$ outliers from their fits are driven by younger ages. Our interpretation of their results is that 
these `outliers' in fact fall along the more meaningful correlation obtained by fitting the full dataset. 

In summary, the conclusions reached in our work disagree with those of B11 mainly through our preferred approaches for fitting 
and interpreting the data.

\subsection{The origin of the FUV upturn}

One motivation for this work was to explore the age and metallicity dependence of the FUV upturn, as a method to distinguish between the 
`metal-rich single-star hypothesis' (e.g. Yi et al. 1997b) and the `binary-star hypothesis' (Han et al. 2007) for the 
origin of the FUV flux in passive galaxies.  Although only $\sim$13 per cent of 
galaxies in our sample have classic FUV upturns, in the sense of having rising $f_\lambda$-vs-$\lambda$ below 2000\,\AA\ ($FUV-NUV<0.9$), we prefer 
a `continuous' treatment of the colours, rather than imposing a discrete classification as done for instance by Yi et al. (2011). 

In both the Han et al. and Yi et al. models, the FUV upturn is caused by helium-burning stars with very thin hydrogen 
envelopes, having effective temperatures $\sim$25\,000\,K. In addition to this extreme horizontal branch 
(EHB\footnote{Han et al. (2007) refer to these stars more generally as hot sub-dwarfs, since they consider a wide range in formation mechanisms.})
phase itself, the subsequent post-HB evolution of such stars can include long-lived UV-luminous phases 
(e.g. the failed asymptotic giant branch, or AGB-Manqu\'e). 
These stars were identified as probable sources of the UV upturn through their good match to the spectra (both continuum and lines) of strong-upturn 
galaxies like NGC 1399 (e.g. Ferguson et al. 1991). 
\hst\ imaging of M32 by Brown et al. (2000, 2008) has confirmed the presence of EHB and post-EHB stars, and their dominance of the FUV emission
in this (weak-upturn) elliptical. 

The Han et al. and Yi et al. models differ in the channels by which stars reach the EHB, and hence on their dependence on age and metallicity. 
In the single-star model, Yi et al. (following Greggio \& Renzini 1990) postulate EHB formation through an enhanced efficiency of mass loss on the red 
giant branch at high metallicity, leading to low envelope mass and high temperature on the HB. The evolution of such stars after they leave the HB is 
also sensitive to envelope mass, and hence to metallicity, with a larger fraction of high-metallicity stars following the UV-luminous 
AGB-Manqu\'e pathway instead of evolving onto the asymptotic giant branch. Finally, there is a strong age dependence because stars of smaller
initial mass
need to lose less mass to reach the EHB and the UV-bright post-HB phases. 
In the models favoured  by Yi et al. (1997b, their Figure 16), the envelope-mass threshold governing the post-HB evolution 
causes an extremely rapid onset of the UV upturn, e.g. a 2--4\,mag boost in FUV between ages of 
10\,Gyr and 15\,Gyr. For younger ages the evolution is much slower, perhaps equivalent to a slope of $0.3-0.6$\,mag per decade in age. 
The effect of metallicity is primarily to affect the age at which the rapid UV boost occurs, such that metallicity trends at constant age
may be very strong or very weak, depending on the age range considered. 

In the Han et al. scenario, EHB stars form through a variety of binary interactions, including stable Roche-lobe overflow, common-envelope evolution and
mergers of binary  white dwarfs. 
In these models, the FUV-vs-optical colours are only weakly dependent on age for an SSP. For example, from Table 2 of Han et al. (2007), 
fitting the $FUV-r$ colour over a range of 3\,Gyr to 12\,Gyr, we obtain a slope of $-0.3$\,mag per decade in age. 
The influence of metallicity is not fully addressed: although channels for formation of EHB stars are not thought to depend on metallicity, 
their subsequent evolution may do so, and this is not followed in the model. 

In testing the observed age and metallicity dependence of the upturn, we note that the proposed correlations are `indirect': the hot stars responsible for the 
FUV excess are too faint at optical wavelengths to contribute substantially to the spectroscopic indices. Rather, we are testing the hypothesis that (a) the 
mechanisms for {\it producing} these stars are affected by the age and metallicity of the stellar population from which they evolved, and (b) 
that this population was either the same as, or at least correlated with, the stellar population which today dominates the optical spectrum.

We adopt the observational results for $FUV-i$ as the most direct tracer of the upturn population (see Section~\ref{sec:fuvnuv} for comments on $FUV-NUV$).
The fits presented in Section~\ref{sec:trends} suggest that $FUV-i$ is primarily sensitive to age, becoming bluer for older galaxies with a 
slope of 1.1$\pm$0.1\,mag per decade in age. At fixed age, there is also a dependence on metallicity, such that $FUV-i$ becomes bluer by 
$1.1\pm0.2$\,mag per decade increase of Mg/H. 
The sense of the observed age trend is as predicted by both the Han et al. and Yi et al. models, but its slope is much steeper than predicted by
Han et al. As noted above, the age dependence in Yi et al. is non-linear, becoming extremely steep at ages above 10\,Gyr. 
Some steepening is evident in the upper left panel of Figure~\ref{fig:cols_vs_pops} as an increased scatter
on the blue side of the distribution at ages above $\sim$10\,Gyr. Quantifying this by splitting the sample at the median age of 9\,Gyr, we find that
the older galaxies show a slope of $1.5\pm0.3$ per decade in age, while the younger galaxies have only a marginal dependence $0.5\pm0.2$. 
This behaviour is difficult to reconcile with the Han et al. model, but is a generic prediction of the Yi et al. scenario. 
We note that a strong dependence at the oldest ages is also suggested by the apparently rapid redshift evolution of the UV upturn (e.g. Brown et al. 2003; Ree et al. 2007, but
see also Atlee, Assef \& Kochanek 2009). 
The Yi et al. model also predicts that the FUV-to-optical colour becomes bluer with increasing metallicity, but the strength of the metallicity effect 
itself depends on age. Based on figure 16 of Yi et al. (1997a), we would expect a trend of order 10\,mag per decade in Mg/H for galaxies with ages of 15\,Gyr, 
but only $\sim$0.8\,mag per decade at 5--10 Gyr ages. The latter is broadly consistent with the trend derived from our fits.

Despite recovering a strong age dependence in the $FUV-i$ colour, we continue to find a large scatter around the average trend, with 
variations in the optically-dominant stellar populations accounting for only $\sim$39 per cent of the total variance in colour,
and a further 19 per cent is contributed by measurement error. There are only marginal correlations of $FUV-i$ with element abundance ratios, which do not substantially reduce the scatter. Hence a simple, deterministic, explanation of the FUV colours appears inconsistent with SSP models, at least to the extent that they can be constrained from the optical spectra. 
Considering more realistic extended star-formation histories (SFH), in combination with the threshold age effect in the Yi et al. models, 
perhaps offers a more promising explanation for the scatter in $FUV-i$. 
The SSP-equivalent age is essentially a weighted average over the true distribution of stellar ages.
Because in the Yi et al. scenario the oldest populations produce dramatically more FUV flux, 
the $FUV-i$ colours for a galaxy with extended SFH will be sensitive to the  fraction of stars in the old tail of the age distribution. 
As a simplified example, consider a case where populations older than 12\,Gyr receive an instantaneous 2\,mag boost to their FUV flux. 
A galaxy with gaussian SFH with mean age $t_0$=10\,Gyr and dispersion $\sigma_t$=1\,Gyr has 2.3 per cent of its stellar content FUV-boosted, 
resulting in a total flux  0.12\,mag brighter in FUV than an SSP of the same age. Keeping the same mean age, but increasing the dispersion 
to $\sigma_t$=1.5\,Gyr results in boosting $\sim$9.1 per cent of the stellar content, increasing the total flux by a {\it further} 0.31\,mag\footnote{
In general, for a gaussian SFH with mean age $t_0$, age dispersion $\sigma_t$, and a flux-boost factor $b$ imposed on ages above $T$, the 
total flux is increased by a factor $f = \frac{1}{2}(1+b)+\frac{1}{2}(1-b)\ {\rm erf}\left(\frac{T-t_0}{\sqrt{2}\sigma_t}\right)$.
}.
Thus if the FUV luminosity increase for the oldest galaxies is sufficiently strong and rapid, then the observed excess scatter of $\sim$0.3 in 
$FUV-i$ could be attributable to a modest variation in star-formation time-scale, at a given mean age\footnote{
However, if the $\alpha$ element abundance ratio is interpreted as a star-formation time-scale indicator (Thomas et al. 2005), then this
explanation also predicts that bluer $FUV-i$ (at given age) would be accompanied by higher Fe/Mg ratios, contrary to the observed trend.}. 
Further investigation is required to test whether this explanation is viable given more realistic and detailed models of the flux evolution. 

In conclusion, the observed relationship between $FUV-i$ colour and spectroscopic age and metallicity appears to favour the 
metal-rich single-star EHB hypothesis for the origin of the UV upturn. This scenario generically predicts a dependence on age
in particular among the oldest galaxies, and on metallicity. The large scatter at given SSP-equivalent age and metallicity can possibly be
explained through galaxy-to-galaxy variations in early star-formation history. The binary hypothesis cannot account naturally for these results.

\subsection{Excess scatter at NUV:  residual star-formation, or UV-upturn leakage?}\label{sec:leakage}

A second motivation for our study was to investigate the cause of the large scatter seen in NUV-vs-optical colours. 

The NUV-to-optical scatter has been interpreted as evidence for recent or widespread ongoing star formation in optically-red galaxies (e.g. Kaviraj et al. 2007). 
Sample definition is critically important to this result, since objects with {\it ongoing} star formation will evidently have blue UV colours. 
For our analysis, we excluded galaxies with central  H$\alpha$ emission from the spectroscopy. 
The remaining objects have only $\sim$0.16\,mag scatter around the $NUV-i$ colour--magnitude (or colour$-\sigma$) relation, compared to values in the 
range 0.3--0.6\,mag reported
in the literature (Yi et al. 2005; Boselli et al. 2005; Haines, Gargiulo \& Merluzzi 2008;  Rawle et al. 2008). 
Star formation occurring further from the nucleus cannot be excluded, but does not appear to dominate the colour scatter (Section~\ref{sec:mismatch}). 

We now consider the correlations of $NUV-i$ with spectroscopic age and metallicity. 
Unlike in the FUV case we are here testing `direct' correlations, in the sense that the {\it same} stars (in particular warm main-sequence turn-off stars) are expected
to contribute to the NUV colour variations and to the spectroscopic ages. 
We find significant trends for bluer colours at lower metallicity (the dominant effect, as found also by Rawle et al. 2008) and at younger age. 
While the age dependence is  qualitatively as expected from evolutionary synthesis models, the slopes that we obtain from the
 fits (e.g. $\sim$0.2\,mag per decade in age) are much shallower than the predictions derived by Dorman et al. (2003), 
 who report a 2--3\,mag change in $NUV-V$ per decade in age. The evolution of $NUV-i$ in the Bruzual \& Charlot (2003) models is 
 more gradual, with $\sim$0.65\,mag change per decade in age, though still steeper than  our measured trends. 
(The Bruzual \& Charlot models do not include blue horizontal branch stars; their presence would further dilute the predicted age trend.)

Age and metallicity variations account for only 20\,per cent of the total variation in $NUV-i$, leaving a residual scatter of 0.15\,mag around the fit. 
Splitting the sample by spectroscopic age, we find that the fit is poorer (i.e. larger scatter and smaller $R^2$) for the {\it oldest} galaxies in the sample.
For galaxies older than the median age of 9\,Gyr, the age and metallicity effects account for only 12\,per cent of the variance. For galaxies younger than 
9\,Gyr, correlations with age and Mg/H jointly account for 44\,per cent of the total variance.

As noted elsewhere,  the SSP-equivalent is only a weighted average over the unknown true SFH of the galaxy. For example,  an SSP age of $\sim$5\,Gyr
may represent  a population formed mostly at early epochs, but with a small mass-fraction added in the past $\sim$Gyr. 
We have tested whether the remaining scatter in $NUV-i$ is due to young sub-populations, using the H$\delta_F$ index. 
Discrepancies in H$\delta_F$ compared to the (H$\beta$-based) SSP-equivalent age estimates can be interpreted as evidence for star-formation within
the past $\sim$2.5\,Gyr (Serra \& Trager 2007), or alternatively as evidence for contributions from other warm stars, e.g. on the blue HB 
(Schiavon et al. 2004). We consider the residuals from the fit of $NUV-i$ versus age and metallicity, and compare them to the residuals from a fit 
of H$\delta_F$ against age and metallicity (third panel of Figure~\ref{fig:resids}. The residuals from the two fits are anti-correlated, as expected: 
galaxies with stronger H$\delta_F$ than expected from their SSP-equivalent ages and metallicity 
are also bluer than the average galaxy at that age and metallicity.
The correlation is significant at the $\sim$4$\sigma$ level, and this dependence
accounts for $\sim$8\,per cent of the total variance in $NUV-i$. Thus, variations in true star-formation history (or blue HB content), as traced by H$\delta_F$ make
a measurable, but minor, contribution to the remaining $NUV-i$ scatter at a given SSP-equivalent age and metallicity. 
(Equivalent tests for the other colours reveal 2--3$\sigma$ trends for $u^*-g$ and $g-i$, which $\la$3\,per cent to the variance in these colours.)

The above results suggest that the excess scatter at NUV is related to old populations, rather than to young stars. 
We hence explore here whether $NUV-i$ is significantly affected by the old helium-burning stars, i.e. the UV-upturn sources. 
The possibility that such sources make a significant contribution to the flux at NUV has previously been discussed
in previous work (e.g. Dorman et al. 2003; Rawle et al. 2008).
If we adopt $FUV-i$ as an indicator for the strength of the upturn population, we find that $\Delta(NUV-i)$,
defined as the residual from the age--metallicity fit for this colour, is correlated with $FUV-i$ at the 6$\sigma$ level (fourth panel of Figure~\ref{fig:resids}).
(By contrast the $NUV-i$ colour itself shows only a $\sim$2$\sigma$ correlation with $FUV-i$, because other factors {\it are} important in driving the  NUV flux.) 
This result suggests that some of the excess scatter in the fits to $NUV-i$ results from `leakage' of the FUV upturn.
Incorporating $FUV-i$ as an additional term in a simultaneous fit for $NUV-i$ reduces the scatter from 0.15\,mag to 0.13\,mag.
In the resulting fit, the coefficient of $FUV-i$ is $0.31\pm0.04$; the slope with $\log T_{\rm SSP}$ is increased to $\sim$0.55
(in better agreement with the Bruzual \& Charlot models),  and that of [Mg/H] is increased to $\sim$1.0.
The resulting coefficient of determination is increased from $R^2=0.20$ to $R^2=0.48$. The effect of `correcting' $NUV-i$ for the FUV leakage can be seen
in the upper-right panel of Figure~\ref{fig:fuvnuvkrig}. Comparison to the equivalent panel for uncorrected $NUV-i$ shows that the
correction makes the oldest galaxies redder (since these are the galaxies with strongest UV upturn) and hence yields a map similar in form to the 
optical colours. 
It has been suggested that the FUV-upturn sources may even affect the `optical' U band (e.g. Yi et al. 1997b). 
However, we find no correlation of the $u^*-g$ (or $u^*-i$) residuals with $FUV-i$, equivalent to the trends seen at NUV. 
We infer that ``leakage'' of the upturn does not extend  significantly to the fairly red $u^*$ bandpass used in our MegaCam observations. 

Thus, whatever the sources of the FUV variation might be, they are likely to cause a significant fraction of the scatter
observed in the NUV as well. Of course, this result does not, on its own, prove that the correlated excess variation in FUV and NUV are both due to 
old hot stars (i.e. the classic UV upturn population), rather than both being due to residual ongoing star formation. We consider the latter possibility unlikely though,
because on other grounds it is clear that the UV upturn itself is {\it in general} not due to young stars (O'Connell 1999, and references therein). 
Moreover, recall that $FUV-i$ is bluer for older spectroscopic age. If ongoing star formation were responsible for variations in $FUV-i$, it
would have to occur preferentially in galaxies that have {\it not} formed stars at the intermediate ages ($\sim$1\,Gyr) which strongly affect the SSP fits.

Finally we note that Salim \& Rich (2010) have presented \hst\ FUV imaging revealing clear signatures of star-formation in a sample of
`quiescent early-type galaxies'  that have no detectable H$\alpha$ in SDSS\footnote{Although in fact half of the examples they show are 
evidently spirals on the basis of the SDSS imaging alone.}. The Salim \& Rich sample however was constructed explicitly to select galaxies that 
are {\it unusually} blue in $FUV-i$, relative to their other properties, so that by definition they are outliers from the overall population.
Indeed, almost all of their sample galaxies are much bluer, in both $FUV-i$ and $NUV-i$, than any of our Coma galaxies 
(measured relative to the peak of the red sequence in the respective studies). Although their result confirms that nuclear
H$\alpha$ measurements alone cannot exclude all star-forming galaxies, it does not imply that such galaxies are common
among samples selected to be {\it representative} of the optical red/passive sequence. 

We conclude that a majority of the observed scatter in the NUV-vs-optical colour can be adequately accounted for by a combination of (i) variation in age and metallicity as inferred from optical spectroscopy ($\sim$20 per cent of total variance), (ii) a significant contribution from the same old hot stars which dominate the FUV scatter ($\sim$30 per cent), (iii) a minor contributions from abundance ratio effects and variation in SFH at given SSP-equivalent age ($\sim$15\,per cent) and (iv) measurement error ($\sim$15 per cent).
There is little need to invoke widespread `residual star formation' to explain the NUV colour scatter in our sample.

\begin{table*}
\caption{Summary of the sources of variance in the colours as determined in this paper. The contributions are expressed as a fraction (in per cent)
of the total variance in each colour. We distinguish `astrophysical' sources of variance, i.e. correlations with intrinsic  galaxy properties, from
 those due to systematic errors in the measurements. 
The sum of these constitute the total `structural' component of the variance, i.e. the fraction attributable to known dependencies. 
Adding the variance due to random errors, we arrive at the total fraction of variance that has been accounted for. 
All components below the $\sim$4\,per cent level are omitted for clarity. }\label{tab:summary}
\begin{tabular}{llcccccc}
\hline
Contribution & 
	& \ \ $FUV-i$ \ \ 
	& \ \ $NUV-i$ \ \ 
	& \ \ $FUV-NUV$ \ \ 
	& \ \ $u^*-g$ \ \ 
	& \ \ $g-i$ \ \ & Section \\
\hline
Astrophysical:\\
&Age/metallicity      & {\bf 39} & {\bf 20} & {\bf 62}  & {\bf 44} & {\bf 63} & 3.3\\
&Abundance ratios  & 5 & 8 & 4 & 6 & 8 & 3.5\\
&Young sub-populations		        & & 8 & &  &  & 4.3\\
& UV-upturn leakage & & 28 & & & & 4.3\\ 
Systematics:\\
&Photometric calib  & &  & & 19 & & 3.4.1\\
&$k$-correction  	& & & & 6 & & 3.4.2\\
~\\
Total structural component & & {\bf 44} & {\bf 64} & {\bf 66} & {\bf 75} &  {\bf 72} \\
Random errors                     & & 19 & 17 & 21 & 9 &  9 & 3.1 \\
Total explained	 component	  & & {\bf 63} & {\bf 81} & {\bf 87} & {\bf 83} & {\bf 81} \\
\hline
\end{tabular}
\end{table*}

\subsection{A comment on the tight correlations for FUV--NUV}\label{sec:fuvnuv}

We have shown that the overall strength of the UV upturn, as traced by $FUV-i$, shows scatter that cannot be accounted for by 
variations in the optically-dominant stellar populations.
However, the $FUV-NUV$ colour is surprisingly well behaved, in the sense that most of its variation is predictable from the velocity dispersion, or from age and metallicity, 
with a precision similar to that found for the optical colours. (This statement is in fact equivalent to our conclusion that leakage from the UV upturn 
contributes much of the excess scatter in $NUV-i$.)  
Age and metallicity variations account for $\sim$60 per cent of the total variance in  $FUV-NUV$, with a further $\sim$20 per cent attributable to measurement errors. 

That the $FUV-NUV$ colour follows tighter relations than the FUV-vs-optical colours has been noted before (Donas et al. 2007), 
and B11 conclude that $FUV-NUV$ should be the preferred colour to investigate the UV upturn. 
We disagree with this position: given the consensus that the upturn sources are hot old stars, the crucial quantity is their {\it incidence} relative to cool stars of 
similar age, which is best traced using the red-optical or near-infrared flux. 

Simplistically, the $FUV-NUV$ colour should be sensitive instead to the {\it temperature} of the upturn sources, which would provide further constraints on the 
origin of the upturn. For instance, Figure 14 of Yi et al. (1997b) shows that within their model, the post-EHB stars generate more FUV flux than the EHB itself, 
while in the NUV the contributions of EHB and post-EHB  are comparable. 
In practice however, any attempt to extract information from the $FUV-NUV$ colour alone will be confused by its sensitivity to at least three influences: 
(i) the age and metallicity of the pre-HB stellar population, via the main-sequence turn-off, 
(ii) the overall strength of the UV-upturn population, and
(iii) the spectral slope of the upturn. 
In particular the strong observed $FUV-NUV$ correlations likely arise from a compounding of effects (i) and (ii): increasing age and metallicity have 
the effect of suppressing the NUV flux from the main sequence, as well as boosting the FUV flux from the helium-burning populations.

We conclude that the $FUV-NUV$ colour will be of limited use in deciphering the origin of the upturn, unless additional information
from longer wavelengths is incorporated.

\section{Conclusions}\label{sec:concs}

We have analysed the UV and optical colours of 150 galaxies in the Coma cluster, selected to lie on the 
(optical) red-sequence and to have no detectable H$\alpha$ emission. Using ages and metallicities derived 
from optical spectroscopy, we have performed a purely empirical test for the origin of scatter in the colours. 
Our primary conclusions are as follows: 

\begin{enumerate}

\item
All of the UV and optical colours we investigate show strong ($>5\sigma$) correlations with luminosity and with velocity dispersion.

\item
The average trends in galaxy colours as a function of velocity dispersion can be accounted for by their correlation with 
the spectroscopically-measured SSP-equivalent ages and metallicities. $NUV-i$, $u^*-g$ and $g-i$ become redder with 
increasing age or metallicity, while $FUV-NUV$ and $FUV-i$ become bluer.

\item
For the optical colours, and for $FUV-NUV$, most (70--80 per cent) of the variance in galaxy colours can be accounted for by correlations with spectroscopic
age and metallicity, once random and systematic error sources are included.

\item
For the $FUV-i$ colour, which traces the UV upturn, a strong anti-correlation with age is observed, combined with 
an anti-correlation with metallicity. The correlations are in the sense predicted by single-star EHB models for the origin of the FUV flux in 
passive galaxies (Yi et al. 1997b), and contrary to the weak dependences predicted by the binary hypothesis of Han et al.  (2007).

\item 
However, only $\sim$60 per cent of the variance can be accounted for, even when measurement errors are included. 
Moreover the galaxies with the oldest SSP-equivalent ages ($>10$\,Gyr) apparently show an excess flux in FUV which is not captured by the linear fits. 
Such a steepening at high age is predicted by the single-star EHB models. The excess scatter at FUV could result from a varying contribution from such ancient stars
in galaxies with younger SSP-ages but extended star-formation histories.

\item
Excess scatter is also observed for the colour $NUV-i$. The residuals are highly correlated with the $FUV-i$ colour, suggesting that the
stars responsible for the UV-upturn also contribute significantly to the scatter in NUV. 
Variation in the true extended star-formation histories, at a given value of the SSP-equivalent age, makes only a small contribution to the $NUV-i$ colour scatter. 
There is no need to invoke widespread `residual star formation' to account for the observed NUV scatter in our sample.

\item
At fixed age and Mg/H, colours are weakly correlated with the Fe/Mg and C/Mg abundance ratios, with no significant dependence on Ca/Mg or N/Mg.
Abundance ratio variations, at fixed Mg/H, contribute only modestly to the scatter in the UV colours. 

\item
Combining all effects addressed in this paper, we can account for 80--90\,per cent of the scatter in all of the colours studied, except for $FUV-i$. 
The contributions are summarised in Table~\ref{tab:summary}.

\end{enumerate}

Because we have limited our analysis to empirical correlations, the above results are independent of any specific population synthesis models for the UV colours. 
The measured correlations and variances should provide a good comparison dataset for future model-based investigations.

\section*{Acknowledgments}
RJS was supported for this work by STFC Rolling Grant PP/C501568/1 ``Extragalactic Astronomy and Cosmology at Durham 2008--2013''.
We are grateful Ann Hornschemeier, Derek Hammer and Tim Rawle for contributions to the GALEX proposal and comments on drafts of the paper, 
and to our other collaborators in the Coma Hectospec survey. We thank Zhanwen Han for comments about the binary star models for the UV upturn, 
and the anonymous referee for numerous helpful suggestions. 
This work is based on observations made with the NASA {\it Galaxy Evolution Explorer (GALEX)}.
 {\it GALEX} is a NASA Small Explorer, developed in cooperation with the Centre National d'Etudes Spatiales of France and the Korean Ministry of Science and Technology. 

Funding for the SDSS and SDSS-II has been provided by the Alfred P. Sloan Foundation, the Participating Institutions, the National Science Foundation, the U.S. Department of Energy, the National Aeronautics and Space Administration, the Japanese Monbukagakusho, the Max Planck Society, and the Higher Education Funding Council for England. The SDSS Web Site is http://www.sdss.org/.
The SDSS is managed by the Astrophysical Research Consortium for the Participating Institutions. The Participating Institutions are the American Museum of Natural History, Astrophysical Institute Potsdam, University of Basel, University of Cambridge, Case Western Reserve University, University of Chicago, Drexel University, Fermilab, the Institute for Advanced Study, the Japan Participation Group, Johns Hopkins University, the Joint Institute for Nuclear Astrophysics, the Kavli Institute for Particle Astrophysics and Cosmology, the Korean Scientist Group, the Chinese Academy of Sciences (LAMOST), Los Alamos National Laboratory, the Max-Planck-Institute for Astronomy (MPIA), the Max-Planck-Institute for Astrophysics (MPA), New Mexico State University, Ohio State University, University of Pittsburgh, University of Portsmouth, Princeton University, the United States Naval Observatory, and the University of Washington.
\label{lastpage}

{}

\end{document}